\documentclass[iop,numberedappendix]{emulateapj}
\slugcomment{Accepted by \apj} 
\usepackage{natbib}
\usepackage{amsmath}
\usepackage{booktabs}
\usepackage{longtable}
\usepackage{color}
\usepackage{graphicx, subfigure}
\usepackage{hyperref}
\usepackage{multirow}
\usepackage{CJKutf8}

\newcommand{\di}{\mathrm{d}} 

\newcommand{\mr}{\mathrm} 
\newcommand{\CO}{\mathrm{CO}} 
\newcommand{\Ht}{\mathrm{H_2}} 
\newcommand{\Ho}{\mathrm{H}} 
\newcommand{\CR}{\mathrm{CR}} 
\newcommand{\FUV}{\mathrm{FUV}} 
\newcommand{\pc}{\mathrm{pc}} 
\newcommand{\Myr}{\mathrm{Myr}}
\newcommand{\cm}{\mathrm{cm}}

\defcitealias{KO2017}{KO2017}
\defcitealias{Gong2018}{GOK2018}

\begin{document}
\title{The environmental dependence of the $X_\CO$ conversion factor}
\author{Munan Gong (\begin{CJK*}{UTF8}{gbsn}龚慕南\end{CJK*})\altaffilmark{1},
    Eve C. Ostriker\altaffilmark{2}, Chang-Goo Kim\altaffilmark{2},
    Jeong-Gyu Kim\altaffilmark{2}}
\altaffiltext{1}{Max-Planck Institute for Extraterrestrial Physics,
Garching by Munich, 85748, Germany; 
munan@mpe.mpg.de}
\altaffiltext{2}{Department of Astrophysical Sciences, Princeton, New Jersey
08544, USA; eco@astro.princeton.edu}

\begin{abstract}
  $\CO$ is the most widely used observational tracer of molecular gas.
  The observable $\CO$
  luminosity is translated to $\Ht$ mass via a conversion
  factor, $X_\CO$, which is a source of uncertainty
  and bias.  Despite variations in $X_\CO$, the
  empirically-determined solar neighborhood value is often applied
  across different galactic environments.  To improve
  understanding of $X_\CO$, we employ
  3D magnetohydrodynamics simulations of the interstellar medium
  (ISM) in galactic disks with a large range of gas surface densities,
  allowing for varying metallicity, far-ultraviolet (FUV) radiation,
  and cosmic ray ionization rate (CRIR).  With the TIGRESS simulation framework
  we model the three-phase ISM with self-consistent star formation and
  feedback, and post-process outputs with chemistry
  and radiation transfer to generate synthetic CO(1--0) and (2--1) maps.
  Our models reproduce the observed
  CO excitation temperatures, line-widths, and line ratios in
  nearby disk galaxies. $X_\CO$ decreases
  with increasing metallicity, with a power-law slope of $-0.8$ for the
  (1--0) line and $-0.5$ for the (2--1) line. $X_\CO$ also decreases
  at higher CRIR, and is insensitive to the FUV radiation. 
  As density increases, $X_\CO$ first decreases due to increasing
  excitation temperature, and then increases when the emission is fully 
  saturated. We provide fits between $X_\CO$ and 
  observable quantities such as the line ratio, peak antenna
  temperature, and line brightness, which probe local gas conditions. These
  fits, which allow for varying beam size,  may be
  used in observations to calibrate out systematic biases.  We also
  provide estimates of the CO-dark $\Ht$ fraction at different gas
  surface densities, observational sensitivities, and beam sizes.
\end{abstract}

\section{Introduction}
Molecular clouds are the cradles for star formation in
galaxies. Measuring the total molecular content as well as the
distribution and properties of molecular clouds is therefore crucial
to empirical characterization of star formation itself and of the
energy returned by massive young stars to the ISM.
Although $\Ht$ is the most abundant molecule in the ISM, it is
difficult to observe in emission due to its low mass and lack of
dipole moment.  As a result, the second most abundant molecule, $\CO$,
is often used to trace $\Ht$. However, $\CO$ emission is usually 
optically thick, and the standard technique relies on applying a
conversion factor to translate the observed $\CO$ line brightness
$W_\CO$ to the column density of molecular hydrogen $N_\Ht$,
\begin{equation}
    X_\CO \equiv \frac{N_\Ht}{W_\CO}.
\end{equation}
Equivalently, the total molecular gas mass
surface density (including helium) is
obtained as $\Sigma_\mr{mol} = \alpha_\CO W_\CO$ using a conversion factor
$\alpha_\CO = 2.8 m_p X_\CO$.

Traditionally, $X_\CO$ is defined for emission in the $J=1-0$ rotational transition (hereafter denoted as (1--0)). 
It can be
measured empirically by determining the $\Ht$ mass using dust
emission or extinction, gamma-ray emission, or the virial theorem 
\citep[e.g.][]{Dame2001, Lombardi2006, SM1996, Solomon1987}. The average value
of $X_\CO$ in the Milky Way solar neighborhood is $X_\mr{CO, MW}=
2\times 10^{20} ~\mr{cm^{-2}K^{-1}km^{-1}s}$, corresponding to
$\alpha_\mr{CO, MW} = 4.3~ M_\odot \mr{pc^{-2}K^{-1}km^{-1}s}$
\citep[see review by][]{BWL2013}. Often, values of $X_\mr{CO,20} \equiv X_\CO/(10^{20}~\mr{cm^{-2}
    K^{-1} km^{-1} s})$ are reported, and we will adopt this shorthand for numerical results. 

Recently, interferometers such as ALMA have enabled high resolution observations
in nearby galaxies, revealing unprecedented details of molecular clouds in
a wide range of environments down to scales of tens of parsecs
\citep[e.g.][]{Schinnerer2013, Leroy2016, Egusa2018,Faesi2018, Sun2018, Sun2020}. 
However, the environmental dependence of $X_\CO$ is not well understood, and
can introduce significant uncertainties and biases
in measuring the mass and pressure
of molecular gas \citep{Sun2020}.
In addition, many observations are conducted using the CO(2--1) line in
order to achieve higher resolution, and often a fixed ratio of the (2--1)/(1--0)
line intensity is adopted in order to estimate $X_\CO$ \citep{Gratier2010, Sun2020}.

The uncertainties in $X_\CO$ stem from the fact that the value of $X_\CO$ is
observed to vary both locally on small scales within individual molecular
clouds where the volume and column density
as well as thermal and turbulent motions vary
\citep[e.g.][]{Solomon1987, Pineda2008, Ripple2013, Kong2015};
and on large scales across galaxies where the total gas surface density and
velocity dispersion as well as 
environmental conditions such as the metallicity and 
gas heating rate are nonuniform 
\citep[e.g.][]{Israel1997, DS1998, Leroy2011, Sandstrom2013, BWL2013}.
To make the most of the new molecular observations, it is essential 
to understand and calibrate the variations in $X_\CO$.

Many efforts have been made to investigate $X_\CO$ using theoretical models. The approach in \citet{WHM2010} combines the 
comprehensive chemical network of a photodissociation region 
(PDR) code with a highly simplified spherical cloud model.
\citet{Accurso2017} further coupled radiation
from stellar populations to similar spherical cloud models. 
These studies both
allow for comprehensive chemical networks, but lack the realistic 
density and velocity structure produced by turbulence in molecular clouds and their environments 
To model more realistic, turbulent  molecular clouds,
several studies have
employed 3D numerical hydrodynamic and magnetohydrodynamic  (MHD) 
simulations to investigate $X_\CO$
\citep[e.g.][]{GM2011, Shetty2011, Shetty2011b,GC2012a, Szucs2016}.
The molecular clouds in these simulations are
modeled in domains with sizes from parsec to tens of parsecs,
and are effectively isolated from the galactic  ISM.
Their physical properties such as the density, cloud size,
and velocity structure are set
by hand via initial conditions and turbulent driving specified in the
simulations, and radiation fields impinging on the cloud must also
be specified by hand. At the other extreme, 
galaxy simulations have also been used to explore variations in $X_\CO$ 
\citep[e.g][]{Narayanan2011, Narayanan2012, Feldmann2012, Duarte-Cabral2015, Li2018}.
These models can  capture global environmental variations, but with
resolutions coarser than tens of parsecs individual
molecular clouds are not
resolved, and sub-grid models are required to estimate the $\CO$ brightness. 
Due to the computational cost limitations, most of these cloud- and galaxy-scale simulations obtain the chemical abundances of $\Ht$ and $\CO$ 
from either sub-grid models that assume a simplified PDR-like structure within each grid cell
or simplified chemistry networks such as those from \citet{NL1997} and \citet{NL1999}.

In our previous work \citep[][hereafter \citetalias{Gong2018}]{Gong2018}, we
investigated $X_\CO$ using local
galactic disk MHD simulations where massive clouds are
formed self-consistently in the three-phase ISM with star formation and
feedback. We modeled the chemical abundances in post-processing with a compact 
network described in \citet{GOW2016}, which included significant improvements over \citet{NL1999} 
and demonstrated good agreement with the comprehensive PDR code in \citet{WHM2010}.
For this study, kpc-scale conditions input to the MHD simulations 
were similar to the 
solar neighborhood environment, and evolution of
the ISM covered more than a full star formation cycle ($\sim 50 \Myr$)
at pc-scale resolution  \citep[][hereafter \citetalias{KO2017}]{KO2017}.
This study demonstrated that a mean
$X_\CO \approx 0.7-2 \times 10^{20} ~\mr{cm^{-2}K^{-1}km^{-1}s}$ is obtained
(varying somewhat in time and increasing
for large beams), in agreement with Milky Way
observations.  It also showed that $W_\CO$ is sensitive to density, since
collisions are what determines the excitation of rotational transitions.  
Starting from similar local galactic disk models with solar neighborhood-like parameters \citep{Walch2015},
\citet{Seifried2017, Seifried2020} performed zoom-in simulations
of giant molecular clouds (GMCs) with time-dependent chemistry using the \citet{NL1997} network, and achieved a 
resolution of 0.1 pc. They obtained  typical
$X_\CO \approx 1.5 \times 10^{20} ~\mr{cm^{-2}K^{-1}km^{-1}s}$ for a few GMCs,
again in
agreement with observations. Both of these recent studies emphasized that
$X_\CO$ has considerable scatter on small scales.  
Local-box simulations of this kind are particularly advantageous for
investigating $X_\CO$, because they include enough physics to produce
a realistic ISM, while also having high resolution.  However, to date
only solar neighborhood conditions have been considered, not yet
addressing potentially important environmentally-driven variations in
$X_\CO$, such as the dependence on metallicity \citep{BWL2013}.  
Moreover, theoretical models so far have mostly focused on the CO(1--0) line,
although the (2--1) line has been used increasingly in  observations
\citep[e.g.][]{Sun2018}.

In this paper, we build upon \citetalias{Gong2018} to study and
calibrate $X_\CO$ more comprehensively, covering a range of ISM
conditions that prevail in local-Universe galaxies.  As before, we
perform 3D MHD simulations of kpc-sized regions of galactic disks with
$\sim$pc resolution, which produces clouds with realistic density and
velocity structure as determined by self-gravity and turbulence driven
naturally by star formation feedback. The $\Ht$ and $\CO$ abundances
and CO(1--0) and (2--1) line emission maps are obtained via
chemistry and radiation transfer post-processing. By varying the
initial large-scale surface density in the MHD simulations,  as well as the
metallicity, the far ultraviolet (FUV) radiation field strength, and the
cosmic ray ionization rate (CRIR) in the post-processing, we
systematically investigate the dependence of $X_\CO$ on these
environmental parameters. We also study the effect of beam sizes in
our synthetic observations.   We analyze how and why $X_\CO$ depends on
large-scale and small-scale environmental conditions.  We also quantify
the dependence of $X_\CO$ on direct observables (total  CO(1--0) and (2--1)
line strength, peak antenna temperature, and line ratio) that probe gas conditions
for  different models, at a range of observational beam sizes.  Based on the
correlations we identify, we provide 
formulae to calibrate $X_\CO$; these calibrations
can be used to reduce
systematic biases that enter if a constant $X_\CO$ is adopted to convert
observed $W_\CO$ to $N_\Ht$.
The present work may be seen as a natural extension of \citetalias{Gong2018}
beyond solar neighborhood environments.

The structure of this paper is as follows. In Section \ref{section:theory},
we use simple theoretical models to explain the physics 
that enters in setting $X_\CO$;
this provides insight into the environmental dependencies
that may be expected. In Section
\ref{section:method}, we describe the methods adopted for
our numerical MHD simulations, and the post-processing chemistry and
radiative transfer that we use to produce synthetic observations. 
Our results are presented in Section
\ref{section:results}:
first, we describe the overall properties of the simulations in Section \ref{section:overall_properties}; then we validate our simulations by comparing with observations in Section \ref{section:comparison_obs}; Section \ref{section:results_XCO} investigates the dependence of $X_\CO$ on environmental and observable parameters and provides calibration formulae for $X_\CO$; lastly, Section \ref{section:fdark} quantifies the variations in the CO-dark $\Ht$ fraction.
Finally, we summarize our conclusions in Section \ref{section:conclusions}.

\section{Theoretical Expectations}\label{section:theory}
Although the definition of $X_\CO=N_\Ht/W_\CO$ is simple, both $N_\Ht$ and
$W_\CO$ have complex dependencies on many physical parameters. For example, the
cloud density structure influences where both $\CO$ and $\Ht$ form.
The gas kinetic temperature affects collision rates and  hence the
population  of  $\CO$ rotational energy  levels  and transition rates. The
velocity structure affects how much $\CO$ emission can escape the optically
thick dense gas and thus the brightness of the $\CO$ line. The metallicity
changes the formation rate of $\Ht$ and amount of dust shielding available. 
The external FUV radiation 
and CR ionization hinder formation of molecules, while also setting the
gas heating rate. Due to these complex factors,
it is difficult to make an accurate analytical prediction of $X_\CO$ as a
simple function of the environmental variables. However,
reference to simple models is still quite  useful for providing insights into
what $X_\CO$ may depend on, and in which direction. 

Typically, $\CO$ line profiles are not too far from Gaussian, and
to the first order, $W_\CO \propto \sigma_v T_\mr{peak}$, where $\sigma_v$ is
the width of the line and $T_\mr{peak}$ is the peak antenna temperature. From
Section 3.1.2 in \citetalias{Gong2018}, for a uniform slab with optically thick
$\CO$ emission and $T_\mr{peak} \gtrsim 5.5~\mr{K}$, $T_\mr{peak} \approx
T_\mr{exc}$ where $T_\mr{exc}$ is the excitation temperature of the line. Thus,
we can approximate $X_\CO$ as
\begin{equation}\label{eq:XCO_theory}
    X_\CO \equiv \frac{N_\Ht}{W_\CO} \sim \frac{N_\Ht}{\sigma_v T_\mr{exc}}
           \sim \frac{N_\Ht/n}{\sigma_v(T_\mr{exc}/n)}.
\end{equation}
where $n$ is the number density of hydrogen atoms.
The factor in the numerator, $N_\Ht/n$ is determined by the $\Ht$ formation
chemistry, and by the turbulent structure of the molecular clouds.
\citetalias{Gong2018} pointed out that
$\sigma_v$ does not vary as much as $T_\mr{exc}$, so in the denominator the
factor $T_\mr{exc}/n$ is more important for $X_\CO$.

We can make the further assumption that the
molecular gas is either (1) in clouds in approximate 
virial equilibrium with mean density $\rho$ and size
$L_\mr{cloud}\sim \sigma_v/\sqrt{G\rho} $, or
(2) dominating the mass in the  galactic  midplane of an ISM disk that is in
vertical equilibrium,  with scale height $H \sim \sigma_v^2/(G\Sigma_\Ht)$.
In either case, $N_\Ht \propto \sigma_v \sqrt{n}$,
which gives
\begin{equation}\label{eq:XCO_virial}
    X_\CO \propto \frac{\sqrt{n}}{T_\mr{exc}}.
\end{equation}

Taking the CO(1--0) as an example and using a
simplified two-level system model,
\begin{equation}\label{eq:Texc}
    \frac{1}{T_\mr{exc}} = \frac{1}{T_\mr{gas}} + \frac{1}{T_0}\ln \left(
    1 + \frac{\beta A_{10}}{n_c k_{10}} \right)\approx  \frac{1}{T_\mr{gas}}  +
    \frac{\beta A_{10}}{n_c k_{10} T_0}
\end{equation}
from Equation (30) in \citetalias{Gong2018}.  Here, $T_\mr{gas}$ is the gas
temperature,  $n_c$ is the density of the collisional partner ($\Ht$ in
this case), $\beta=(1-e^{-\tau})/\tau$ is the escape probability of the line, $\tau$ is the optical depth of the line,
$k_{10}\approx 6\times 10^{-11} (T_\mr{gas}/100~\mr{K})^{0.2}~\mr{cm^{3}s^{-1}}$
is the 
collisional de-excitation rate, $T_0=5.5~\mr{K}$ characterizes the transition
energy and $A_{10}=7.203\times
10^{-8}~\mr{s^{-1}}$ is the Einstein A coefficient.
If the optical depth $\tau \gg 1$, $\beta \approx 1/\tau$. The expansion of the
logarithm is generally valid for the conditions in molecular clouds, where
$n_c \gtrsim 50\, \cm^{-3}$, $T_\mr{gas} \sim 10-100$K, and $\tau \gtrsim 10$. 

 Using the large velocity gradient (LVG) approximation,
the optical depth is (Equation (7) in \citetalias{Gong2018})
\begin{equation}\label{eq:tau_LVG}
    \tau_\mr{LVG} = \frac{\lambda_{10}^3}{8\pi} \frac{A_{10}n_\CO}{
    |\di v/\di r|}
    f_1 \left( \frac{f_0/g_0}{f_1/g_1} - 1 \right),
\end{equation}
where $\lambda_{10} = 2.6\,{\rm mm}$, $n_\CO$ is the number density of $\CO$
molecules, $g_0=1$ and $g_1=3$ are the degeneracies for $J=0$ and $J=1$ levels, 
$f_0=n_0/n_\CO$ and $f_1 = n_1/n_\CO$ are  the fractions of $\CO$ molecules in
$J=0$ and $J=1$ levels, $n_0$ and $n_1$ are the level populations, and 
$|\di v/\di r|$ is the velocity gradient. If $T_\mr{exc} \gtrsim T_0$, with
the definition of $T_\mr{exc}\equiv T_0/\ln [(f_0/g_0)/(f_1/g_1)]$ and 
$f_0 + f_1 = 1$, then to the first order of ($T_0/T_\mr{exc}$), 
\begin{equation}
    f_1 \left( \frac{f_0/g_0}{f_1/g_1} - 1 \right) =
    \frac{e^{T_0/T_\mr{exc}} - 1}{1 + \frac{g_0}{g_1}e^{T_0/T_\mr{exc}}}
    \approx \left(1 - \frac{g_0}{g_1}\right) \frac{T_0}{T_\mr{exc}}.
\end{equation}
This then gives
\begin{equation}\label{eq:beta_nc}
\frac{\beta A_{10}}{n_c k_{10} }
\approx \frac{24 \pi}{k_{10} \lambda_{10}^3}\frac{|\di v/\di r|}{n^2   f_\CO}\frac{T_\mr{exc}}{T_0},
\end{equation}
assuming $n_c = n_\Ht \approx 0.5n$ in $\CO$ dominated regions; $f_\CO=n_\CO/n$
is the CO abundance relative to hydrogen. 

We consider two limits from Equations (\ref{eq:XCO_virial}), (\ref{eq:Texc})
and (\ref{eq:beta_nc}). In the first case, we consider relatively low $n$.
In this case, $\beta A_{10}/(n_c k_{10})$ is relatively large
(while still allowing the logarithm to be expanded to lowest order),
and  the second term on the right-hand  side of Equation (\ref{eq:Texc})
dominates. Equation (\ref{eq:Texc}) then
gives $\beta A_{10}/(n_c k_{10})\approx   T_0/T_\mr{exc} $, and when combined
with Equation (\ref{eq:beta_nc}) this yields 
\begin{equation}\label{eq:Texc_approx}
    T_\mr{exc} \propto n \left( \frac{f_\CO}{|\di v/\di r|} \right)^{1/2}.
\end{equation}
Finally, inserting in  Equation (\ref{eq:XCO_virial}) we obtain  for the low-density
limit
\begin{equation}\label{eq:XCO_approx1}
    X_\CO \propto \left( \frac{|\di v/\di r|}{n f_\CO} \right)^{1/2}.
\end{equation}
We find that in the simulations, $|\di v/\di r|$ has no systematic density dependence. In this case, as density and $f_\CO$ increase, $X_\CO$ decreases.

The second case we consider is when $n$ is
large, so the first term in the denominator of Equation (\ref{eq:Texc})
dominates. This is the LTE limit of $T_\mr{exc} \to T_\mr{gas}$.
In this high density limit we then  have
\begin{equation}\label{eq:XCO_approx2}
    X_\CO \propto \frac{\sqrt{n}}{T_\mr{gas}},
\end{equation}
which increases with density.  Although $T_\mr{gas}$ does not vary much
within individual dense molecular clouds, it may be higher in environments
with high star formation rates (SFRs) and hence high cosmic ray heating.

We note that the dependencies of $X_\CO$ for low- and high-density limits in
Equations
\ref{eq:XCO_approx1} and \ref{eq:XCO_approx2} are derived using 
over-simplified assumptions,
and thus are never strictly true in realistic molecular
clouds. However, they provide theoretical insight to the behavior that
emerges from much more complex  numerical simulations.
In particular, the above arguments show that $X_\CO$ is not expected to be 
constant on small scales. In fact, we expect $X_\CO$ to have a non-monotonic
relation with density.

On large scales, the main external environmental factors we consider 
in this paper are the FUV
radiation field strength, the CRIR, and the metallicity $Z$. From the simple
photodissociation region (PDR) models in \citet{GOW2016} (for example their 
Figures 5 and 6), we expect that FUV radiation destroys both $\Ht$ and $\CO$.
The CRIR, on the other hand, also impedes both $\Ht$ and $\CO$ formation, but
has the additional effect of heating up the molecular gas and raising the
temperature in $\CO$ dominated regions. Therefore, we expect a larger effect on
$X_\CO$ from the CRIR than from the FUV radiation. 
By raising $T_\mr{gas}$, which tends to increase $T_\mr{exc}$ from
Equation \ref{eq:Texc}, $X_\CO$ will be reduced as the CRIR increases. Equation \ref{eq:Texc} also suggests a higher $X_\CO$ at lower metallicity $Z$, where $f_\CO$ decreases due to lower carbon and oxygen abundances and lower shielding.

Another important observational parameter is $f_\mr{dark}$, 
the fraction of $\CO$-dark $\Ht$. This is defined as the fraction of $\Ht$ with
$\CO$ emission below some detection limit $W_\mr{CO,det}$,
\begin{equation}\label{eq:f_dark}
    f_\mr{dark} = \frac{M_\Ht(W_\CO < W_\mr{CO,det})}{M_\mr{H_2, tot}}.
\end{equation}
Evidently, $f_\mr{dark}$ increases with $W_\mr{CO,det}$. We adopt a constant
$W_\mr{CO,det}$ similar to the PHANGS observations in the main part of this
paper (see Section \ref{section:post-process}),
and further discuss the relation between $f_\mr{dark}$ and $W_\mr{CO,det}$ in
Section \ref{section:fdark}.

\section{Methods}\label{section:method}
The methods used here are very similar to those in \citetalias{Gong2018}, but are
extended to apply to environments beyond the solar neighborhood. We
post-process simulations of galactic disks with chemistry to obtain the
distribution of $\Ht$ and $\CO$, and then use a radiation transfer code to
model the $\CO$ line emission from molecular clouds. Below we briefly
describe our methods and refer the readers to \citetalias{Gong2018} for
more extensive descriptions.

\subsection{MHD simulations\label{section:MHD}}
\begin{figure}[htbp]
\centering
\includegraphics[width=0.95\linewidth]{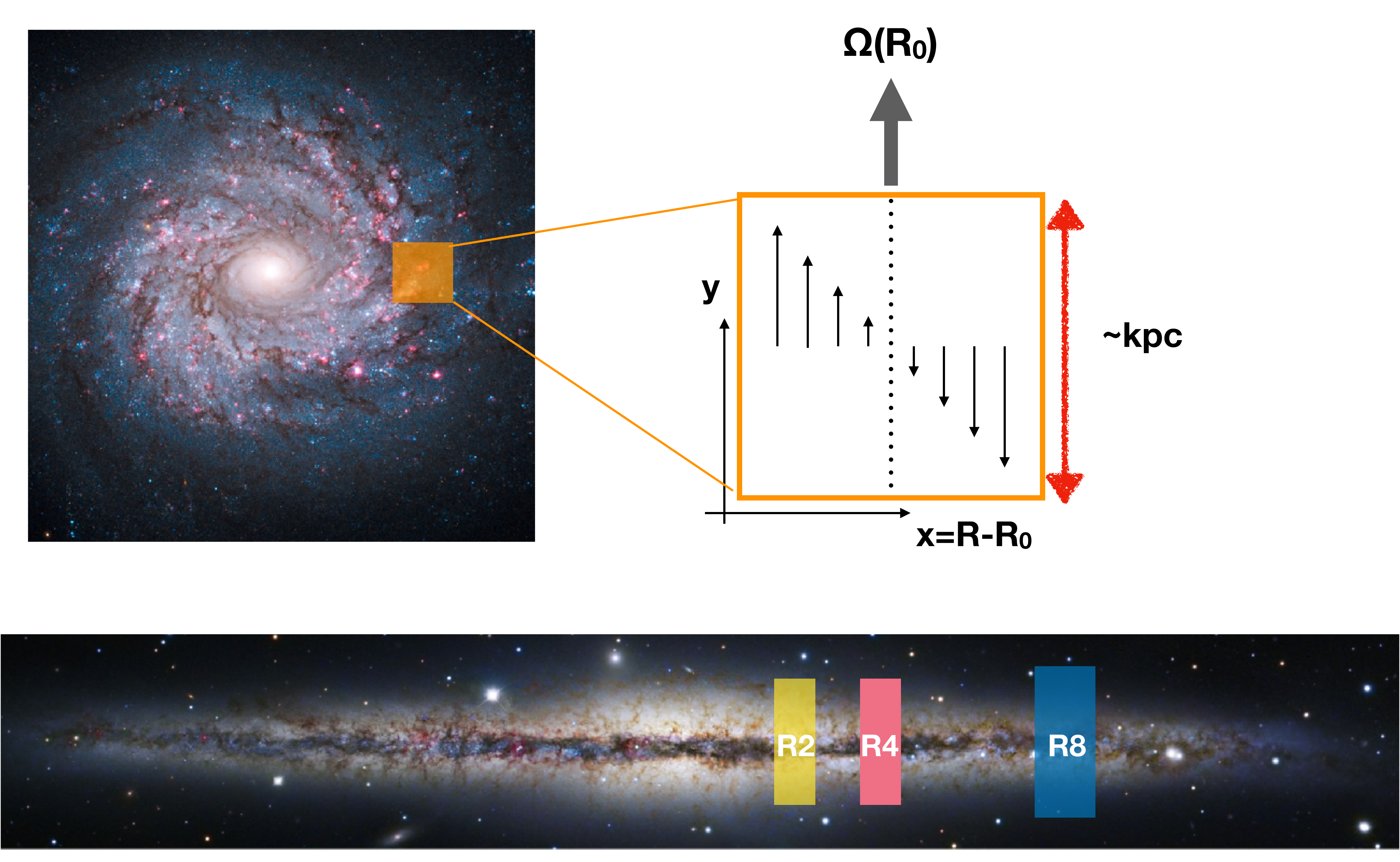}
\caption{Schematic illustration of the TIGRESS framework. R2, R4 and R8 models
  roughly represent the environments in a Milky Way-like
  galaxy at 2, 4, and 8 kpc from the galactic center.
  The gas surface density and SFR decrease from R2 to R4 to R8.
  Image credit: face-on galaxy NGC 3982: ESA/NASA; edge-on galaxy NGC 891: Robert Gendler, NAOJ, HST/NASA, BYU (Michael Joner, David Laney).
}
\label{fig:schematic}
\end{figure}
The MHD simulation is carried out with the TIGRESS
(Three-phase Interstellar medium in Galaxies Resolving Evolution with 
Star formation and Supernova feedback) framework described by
\citetalias{KO2017}
. 
A schematic illustration of the TIGRESS framework is shown in Figure
\ref{fig:schematic}. Each simulation represents a $\sim$  kpc-sized patch
of a galactic disk where the multiphase ISM is self-consistently modeled
with resolved star formation and feedback.
The simulations are conducted using the {\sl Athena} code \citep{Stone2008,SG2009}, 
in a vertically-stratified local shearing box \citep[e.g.][]{SG2010}.
The ideal MHD equations are solved, including gravitational forces from gas,
stars, and the dark matter halo (the 
old stellar disk and the dark matter halo are treated via fixed potentials).
Sink particles are implemented to represent
star clusters \citep{GO2013}, and produce radiation and supernova feedback to the ISM from the  massive stars they contain.
Only core-collapse supernovae are included, from both young star clusters and runaway stars that originated from OB binaries in clusters. The rate of SN explosions is adopted from the stellar population synthesis model STARBURST99 \citep{Leitherer1999}. The FUV radiation from massive stars 
uses  the  same stellar population synthesis model and is based on the instantaneous average  luminosity per unit area over the whole simulated domain, with a simple attenuation factor to account for the mean dust optical depth. This average radiation field is used to obtain the mean heating rate in the atomic ISM (without solving the radiative transfer on-the-fly).

Each  TIGRESS simulation is
run for at least 1.5$t_\mathrm{orb}$ (corresponding to several star formation cycles), where $t_\mr{orb}=2\pi/\Omega$ is the local galactic disk orbital time. A turbulent and
magnetized three-phase ISM with realistic properties emerges. Overall,
quasi-steady state is reached,
with periods of enhanced star formation followed by periods of enhanced
feedback; feedback disperses dense gas, which recollects over time due to
gravity and large-scale converging  flows.  No gas is added to the domain,
but gas is continually lost to galactic winds
\citep{KO2018, KimWT2020, KimCG2020} and to star formation, so
the mean gas surface density declines over time in each simulation.
Much of the volume is occupied by hot ionized gas, and
most of the mass resides near the midplane in the warm and cold neutral medium
(WNM and CNM), similar to the observed ISM in the Milky Way and nearby galaxies.
Although molecular gas is not explicitly modeled in the time-dependent
simulations, it is expected to form within the dense and shielded regions
of the CNM. We model the formation of molecular gas by post-processing the
simulations with
chemistry and shielding, which is described in detail 
in Section \ref{section:post-process}.

\begin{table}[htbp]
    \caption{Galactic Environments in Simulations\tablenotemark{a}}
    \label{table:Rn}
    \centering
    \begin{tabular}{c ccccc}
        \tableline
        \tableline
        Environment &$\Sigma_\mr{gas,init}$ &$\Sigma_\mr{gas}$
        &$\Sigma_\mr{star}$  &$\rho_\mr{DM}$  &$\Omega$\\  
        \tableline
        R2   &150   &40--100 &450 &0.08  &0.1 \\
        R4   &50    &20--40  &208 &0.02  &0.05\\
        R8   &12    &9--11   &42  &0.006 &0.03\\
        \tableline
        \tableline
    \end{tabular}
    \tablenotetext{1}{$\Sigma_\mr{gas,init}$ is the initial gas surface
    density in $\mr{M_\odot pc^{-2}}$. $\Sigma_\mr{gas}$ is the gas surface 
    density range after a quasi-steady state is reached, in $\mr{M_\odot pc^{-2}}$.
    $\Sigma_\mr{star}$
    is the old stellar disk surface density in $\mr{M_\odot pc^{-2}}$.
    $\rho_\mr{DM}$ is the mid-plane dark matter density in 
    $\mr{M_\odot pc^{-3}}$. $\Omega$ is the rotation rate about the center of
    the galaxy, 
    in $\mr{km~s^{-1}pc^{-1}}$.}
\end{table}
We extend the solar neighborhood TIGRESS model from \citetalias{KO2017} (as
previously analyzed in \citetalias{Gong2018}) to a wider range
of environments, as listed in Table \ref{table:Rn} \citep[see also][]{KimCG2020}.
Three types of initial conditions are adopted and the
corresponding MHD models are named R2, R4 and R8.
These very  roughly represent 
environments in a generic Milky Way-like galactic disk at
radial distances of 2, 4, and 8 kpc  
from the galactic center (see Figure~\ref{fig:schematic}). All of the densities (gas, stars, and dark matter)
increase from R8 to  R4 to R2,
closer to the notional galactic center.
As a result of both high gas surface density and  the strong vertical gravity
from the stellar disk, the SFR increases from R8 to R4 to R2. For the
R2 and R4 models, feedback
drives stronger outflows than in the R8 model previously studied in
\citetalias{Gong2018}, especially in the initial stage of the
simulation, leading to a larger decrease in the gas surface density in the
steady state compared to the initial values.
We note that because the simulations are local, the
galactocentric radius does not directly enter the model specification.
The suite of models can therefore equally well be thought of as spanning a
range of galactic environments from low to high values of 
$\Sigma_\mathrm{gas}$ and  $\Sigma_\mathrm{star}$, without regard to the position
in a galaxy.   

\begin{table*}[htbp]
    \caption{MHD Simulation and Post-processing Model Parameters\tablenotemark{a}}
    \label{table:model}
    \centering
    \begin{tabular}{l cccc ccc ccc}
        \tableline
        \tableline
        Model ID &Environment &$\Delta x$ &$L_\mathrm{x,y}$ &$t_\mr{pp}$
        &$Z$ &$f_\CR$ & $f_\FUV$ &$\xi_0$ &$\langle \xi \rangle_{M_{CO}}$
        &$\chi_0$\\
        \tableline
        \multicolumn{7}{l}{Physical environment:}\\
        R2-Z1CR10L10    &R2       &2         &256   &40--80     &1   &1    &1  
        &$(1.6\pm 1.1)\times 10^{-14}$ &$(1.1\pm 0.6)\times 10^{-15}$ &$78\pm 55$\\
        R2-Z1L10       &$\cdots$\tablenotemark{b} 
                                &$\cdots$  &$\cdots$ &$\cdots$  &1   &0.1  &1  \\
        R2-Z1CR10      &$\cdots$ &$\cdots$  &$\cdots$ &$\cdots$  &1   &1    &0.1\\
        \textbf{R2-Z1}&$\mathbf{\cdots}$ &$\mathbf{\cdots}$  &$\mathbf{\cdots}$ &$\mathbf{\cdots}$  &\textbf{1}   &\textbf{0.1}  &\textbf{0.1}
        &$\mathbf{(1.6\pm 1.1)\times 10^{-15}}$ &$\mathbf{(1.1\pm 0.6)\times 10^{-16}}$ &$\mathbf{7.8\pm 5.5}$\\
        R2-Z1L01      &$\cdots$ &$\cdots$  &$\cdots$ &$\cdots$  &1   &0.1  &0.01\\
        R2-Z1CR01     &$\cdots$ &$\cdots$  &$\cdots$ &$\cdots$  &1   &0.01 &0.1\\
        R2-Z1CR01L01  &$\cdots$ &$\cdots$  &$\cdots$ &$\cdots$  &1   &0.01 &0.01\\
        R2-Z05        &$\cdots$ &$\cdots$  &$\cdots$ &$\cdots$  &0.5 &0.1  &0.1\\
        R2-Z2         &$\cdots$ &$\cdots$  &$\cdots$ &$\cdots$  &2   &0.1  &0.1\\
        R4-Z1CR10L10    &R4       &2         &512   &50-160     &1   &1    &1  
        &$(5.1\pm 3.3)\times 10^{-15}$ &$(5.0\pm 2.4)\times 10^{-16}$ &$26\pm 16$\\
        R4-Z1L10       &$\cdots$ &$\cdots$  &$\cdots$ &$\cdots$ &1   &0.1  &1  \\
        R4-Z1CR10      &$\cdots$ &$\cdots$  &$\cdots$ &$\cdots$  &1   &1    &0.1\\
        \textbf{R4-Z1} &$\mathbf{\cdots}$ &$\mathbf{\cdots}$  &$\mathbf{\cdots}$ &$\mathbf{\cdots}$  &\textbf{1}   &\textbf{0.1}  &\textbf{0.1}
        &$\mathbf{(5.1\pm 3.3)\times 10^{-16}}$ &$\mathbf{(5.0\pm 2.4)\times 10^{-17}}$ &$\mathbf{2.6\pm 1.6}$\\
        R4-Z1L01      &$\cdots$ &$\cdots$  &$\cdots$ &$\cdots$  &1   &0.1  &0.01\\
        R4-Z1CR01     &$\cdots$ &$\cdots$  &$\cdots$ &$\cdots$  &1   &0.01 &0.1\\
        R4-Z1CR01L01  &$\cdots$ &$\cdots$  &$\cdots$ &$\cdots$  &1   &0.01 &0.01\\
        R4-Z05        &$\cdots$ &$\cdots$  &$\cdots$ &$\cdots$  &0.5 &0.1  &0.1\\
        R4-Z2         &$\cdots$ &$\cdots$  &$\cdots$ &$\cdots$  &2   &0.1  &0.1\\
        \textbf{R8-Z1} &\textbf{R8}       &\textbf{2}         &\textbf{1024}  &\textbf{300--400}     &\textbf{1}   &\textbf{1}    &\textbf{1}  
        &$\mathbf{(4.7\pm 4.0)\times 10^{-16}}$ &$\mathbf{(9.2\pm 5.8)\times 10^{-17}}$ &$\mathbf{2.4\pm 2.0}$\\
        R8-Z05        &$\cdots$ &$\cdots$  &$\cdots$ &$\cdots$  &0.5 &1    &1  \\
        R8-Z2         &$\cdots$ &$\cdots$  &$\cdots$ &$\cdots$  &2   &1    &1  \\
        \tableline
        \multicolumn{7}{l}{Convergence of simulation box-size:}\\
        R2B2-Z1       &R2       &2         &512   &40--60     &1   &0.1  &0.1
        &$(8.2\pm 3.6)\times 10^{-15}$ &$ $$(4.9\pm 1.8)\times 10^{-16}$ &$4.1\pm 1.8$\\
        R2B2-Z05      &$\cdots$ &$\cdots$  &$\cdots$ &$\cdots$  &0.5 &0.1  &0.1\\
        R2B2-Z2       &$\cdots$ &$\cdots$  &$\cdots$ &$\cdots$ &2   &0.1  &0.1\\
        \tableline
        \multicolumn{7}{l}{Convergence of numerical resolution}\\
        R2N2-Z1       &R2       &1         &256   &51--54     &1   &0.1  &0.1
        &$(5.6\pm 1.1)\times 10^{-15}$ &$ $$(5.3\pm 1.2)\times 10^{-16}$ &$2.8\pm 0.5$\\
        R2N2-Z05      &$\cdots$ &$\cdots$  &$\cdots$ &$\cdots$  &0.5 &0.1  &0.1\\
        R2N2-Z2       &$\cdots$ &$\cdots$  &$\cdots$ &$\cdots$  &2   &0.1  &0.1\\
        \tableline
        \tableline
    \end{tabular}
    \tablenotetext{1}{The fiducial post-processing models for R2, R4 and R8 simulations are marked in bold. $\Delta x$ is the numerical resolution in pc.
    $L_\mathrm{x,y}$ is the box-size in $x$ and $y$ directions in pc. $Z$ is
    the metallicity used in post-processing. $t_\mr{pp}$ is the MHD  simulation 
    time interval from which 
    the snapshots for
    post-processing are taken, in Myr.
    $f_\CR$ and $f_\FUV$ are the
    reduction factors of unattenuated CRIR and FUV radiation field used in
    post-processing (see text in Section \ref{section:post-process}). $\xi_0$
    and $\langle \xi \rangle_{M_{CO}}$ are the unattenuated and $\CO$-mass weighted
    average CRIR in $\mathrm{s^{-1}H^{-1}}$ (after $f_\CR$ is applied).
    $\chi_0$ is the unattenuated FUV radiation field intensity in \citet{Draine1978}
    units (after $f_\FUV$ is applied), and $\chi_0=1$ corresponds to $4\pi J_\mr{FUV}=2.7\times 10^{-3}~\mr{erg~cm^{-2}s^{-1}}$. For $\xi_0$, $\langle \xi \rangle_{M_{CO}}$
    and $\chi_0$, the mean values and standard deviations from the simulation
    snapshots used for post-processing are listed.}
    \tablenotetext{2}{``$\cdots$'' represents that the corresponding value
    in the column is the same as the previous row.}
\end{table*}

The physical parameters of the TIGRESS MHD
simulations are summarized as part of Table
\ref{table:model}.
The simulations are conducted using a regular Cartesian grid. Each resolution
element has a size of $\Delta x$ in all three dimensions. The
simulations are run with a resolution of $\Delta x=2~\mr{pc}$. In order to
obtain a higher numerical resolution with limited computational resources, we
restart one of the R2 simulation after it reaches the steady state (at
$50~\mr{Myr}$) with a doubled resolution of $1~\pc$, and run that for $4~\mr{Myr}$.
The boundary condition is shearing-periodic in the $x$ direction, periodic in
the $y$ direction, and outflow in the $z$ direction. 
The simulation box-size is $L_x \times L_y \times L_z$, where $L_x=L_y$ and  
$L_z=3584~\mr{pc}$ for R2 and R4 models and $L_z=7168$ for R8 models. $L_x$ and
$L_y$ increases from $256~\mr{pc}$ in R2 models to $1024~\pc$ in R8 models. 
Larger horizontal box sizes are needed in the lower surface density models, where the expanding bubbles from supernovae explosions are larger due to the lower mean density, and individual superbubbles (created by correlated supernovae explosions) can fill the whole midplane volume if the box size is too small \citep{KimCG2020}. We also carry out a set of R2 models with a larger horizontal 
box size of $L_x = L_y = 512~\mr{pc}$ to investigate the numerical effect of the
changing box sizes.

\subsection{Post-processing $X_\CO$\label{section:post-process}}
To obtain the chemical composition of the gas, we use the
chemistry post-processing module within the code {\sl Athena++} \citep{White2016,
Stone2020} that we developed in \citetalias{Gong2018}.
Because almost all mass and molecular gas resides near the midplane, we
isolate the midplane region of $-512~\mr{pc} < z < 512~\mr{pc}$ for
post-processing.
The code reads the output from the TIGRESS simulations and performs
chemistry calculations assuming the density and velocity in each
grid cell are fixed.

We use the simplified chemical network of \citet{GOW2016},
which gives accurate abundances of $\Ht$ and $\CO$. In order to compute 
the photoionization and photodissociation rates of the chemical species, 
we use the six-ray approximation: in each cell, the radiation field is 
calculated by ray-tracing and averaged over six directions along the Cartesian
axes accounting for the dust and molecular line shielding \citep{NL1997, NL1999, GM2007}. The incident unattenuated radiation field
is assumed to come from the edge of the computational domain along each ray.
The unattenuated FUV radiation is directly obtained from the TIGRESS simulations (see Section \ref{section:MHD}).

The CRIR is similarly calculated with the six-ray method, where $\xi(N_\Ho)$ is computed along each ray and averaged to obtain the final value. We 
adapt the CR attenuation prescription of \citet{NW2017} and \citet{Silsbee2019}, 
\begin{equation}\label{eq:CRatten}
    \xi(N_\Ho) = 
    \begin{cases}
        \xi_0, &N_\Ho \leq N_\mr{H,0} \\
        \xi_0\left( \frac{N_\Ho}{N_\mr{H,0}} \right)^{-1}, &N_\Ho > N_\mr{H,0},
    \end{cases}
\end{equation}
where $N_\mr{H,0} = 9.35\times 10^{20}~\mr{cm^{-2}}$ and $\xi_0$ is the unattenuated CRIR.
We set $\xi_0 = 2\times 10^{-16} \chi_0 \mr{s^{-1}H^{-1}}$, meaning the CRIR is normalized by the cosmic ray rate inferred
from modeling abundances of ions in diffuse molecular clouds  near the Sun \citep{Indriolo2007, NW2017}, 
and proportional to $\chi$,
the unattenuated FUV radiation field intensity in \citet{Draine1978} units 
($\chi_0=1$ corresponds to $4\pi J_\mr{FUV}=2.7\times 10^{-3}~\mr{erg~cm^{-2}s^{-1}}$).
We adopt this approach since both $\xi_0$ and $\chi_0$
are expected to scale roughly with the SFR.

The SFR in the solar
neighborhood model R8 is consistent with observations \citep{KO2017}.
However, the SFR in the R4 and R2 MHD simulations are $\Sigma_\mr{SFR}\approx
0.1-1~\mr{M_\sun yr^{-1} kpc^{-2}}$, about an
order of magnitude higher than the observed values at the corresponding gas
surface density in the nearby disk galaxies \citep{Sun2020}.
In part, this is because the R2 and R4 simulations adopt higher stellar
midplane densities  than are typically found in nearby  galaxies. Stronger
stellar gravity compresses the disk vertically and tends to enhance star
formation.  Additionally,
limitations of the simulations may tend to produce higher-than-realistic
SFR.  One limitation is that only
supernova and FUV radiation feedback were considered in the MHD simulations.
Additional sources of
feedback such as ionizing radiation and stellar wind may play a significant
role in  reality, but were not included in these simulations.  ``Early''
feedback may be particularly important in environments at high density where
gravitational timescales in dense clouds are shorter than the time before
the onset of the first supernova.  We
plan to include these additional feedback mechanisms in the future, and
preliminary results show that SFRs can be decreased by a factor of a few. 
Moreover, the present shearing box simulations do not account for effects of
large-scale galactic structure, such as spiral arms.  Using simulations
that do include spiral structure \citep{KimWT2020}, we have found
that arm regions with $\Sigma_{\rm gas}$ comparable to that in model R4 have
lower local SFR. 
Limited resolution may also tend to produce higher-than-realistic SFRs, since
star cluster particles form instantaneously out of gas at the grid scale
that becomes unresolved (with cluster particle mass $\propto  \Delta x$); at
higher resolution, initial particle masses would be lower and feedback
might be able to prevent accretion of material concentrated near the 
particle.   

To allow for radiation energy input rates that differ from those in the MHD
simulations, we apply reduction factors 
$f_\FUV$ and $f_\CR$ to the unattenuated FUV radiation and CRIR
when we post-process the
simulations to obtain chemical abundances. The fiducial models adopt
$f_\CR=f_\FUV=1$ in R8 and $f_\CR=f_\FUV=0.1$ in R4 and R2 simulations, so that
the corresponding CRIR and FUV radiation in fiducial models
are roughly  in accord with 
observed SFRs at the corresponding surface
densities. We also run a series of models 
varying $f_\CR$ and $f_\FUV$ to
investigate the effect of varying CRIR and FUV radiation on $X_\CO$. Treating
these rates as independent parameters allows us to explore the effects of
heating and dissociation on the CO abundance and excitation.

In post-processing, we also vary the gas and dust metallicity $Z$, which is
defined relative to the metallicity in the solar neighborhood and is the same in
dust and gas. The TIGRESS simulations themselves are conducted assuming
a solar-neighborhood metallicity of $Z=1$, while
we vary $Z=0.5-2$ in the chemistry 
post-processing. Although the treatment is not fully
self-consistent, we will still
capture the effect varying $Z$ on $X_\CO$ better than simple plane-parallel
or spherical models, because the parent MHD models have realistic density
and velocity distributions and correlations.  Varying $Z$ changes
the amount of dust shielding for $\CO$ photo-dissociation \citep{WHM2010},
and also affects the $\CO$ abundance through the abundance of C and O relative
to H input to the chemistry module. 

The physics models for varying post-processing choices are listed in
Table~\ref{table:model}.  Model names encode information regarding the
underlying  MHD  model, the metallicity relative to solar neighborhood, and 
the CRIR and FUV scaling parameters
relative to the fiducial value.  The table also provides values for the
unattenuated CRIR and FUV intensity.  

For chemistry post-processing, we assume an initial chemical composition of hydrogen in the form of $\Ht$ and
all other elements, $\mr{C}$, $\mr{O}$, and $\mr{Si}$, in the atomic form.
The initial number abundances relative to hydrogen are
$x_\mr{C}=1.6\times 10^{-4}Z$, $x_\mr{O}=3.2\times 10^{-4}Z$ and 
$x_\mr{Si}=1.7\times 10^{-6}Z$, following \citet{GOW2016}. The initial temperature
is taken from the output of the TIGRESS simulations.
We evolve the chemistry and temperature simultaneously for time
$t_\mr{chem}=50~\Myr$, so that the chemical abundances and temperature of the gas
reach a steady state.

We use the steady state chemistry and temperature as an
input for the radiation transfer code {\sl RADMC-3D} \citep{Dullemond2012}, to
obtain synthetic observational maps of the CO(1--0) and CO(2--1) 
line emission. We use a passband from -20 to 20 $\mr{km/s}$ (wide enough to include all CO emission), and a velocity resolution of $0.5~\mr{km/s}$.  The velocity gradient $|\di v / \di r|$ is calculated by averaging the absolute velocity gradient across the six faces of each grid cell in the simulation. The total brightness $W_\CO$ is calculated by integrating over all velocity channels. $T_\mr{peak}$ is taken to be the peak antenna temperature over all velocity channels. The velocity dispersion of the line is calculated using $\sigma_v = \sqrt{\langle v^2 \rangle_{T_A} - \langle v \rangle_{T_A}^2}$, where $\langle v \rangle_{T_A} = \int v T_A \di v / \int T_A \di v$ is the antenna temperature (or equivalently, intensity) weighted average of velocity, and similarly  $\langle v^2 \rangle_{T_A} = \int v^2 T_A \di v / \int T_A \di v$.\footnote{Observationally, $\sigma_v$ is often defined as the equivalent width $W_\CO/(\sqrt{2\pi}T_\mr{peak})$, since this definition is less sensitive to noise \citep[e.g.][]{Sun2018}. Because we do not suffer from observational noise, and the line profile is usually close to Gaussian, our moment-based definition gives similar values of $\sigma_v$ to the equivalent width definition.}
The synthetic observations are performed along the z-axis, so that the
observer is looking at the galactic disk face-on. This avoids blending, as
all molecular clouds form near the mid-plane of the galactic disk. 
The default beam size $r_\mr{beam}$ in our synthetic observations is 
the same as the numerical resolution $\Delta x$ in the TIGRESS simulations.
Note that we have a square shaped beam, the same as our numerical resolution
elements\footnote{In \citetalias{Gong2018}, we have compared results for our square beam to the results for a circular Gaussian beam, and find that it makes very little difference for $X_\mr{CO}$.}.
In real observations, the beam size (in
physical units) varies depending on the telescope and the distance of the
object. To investigate the effect of changing $r_\mr{beam}$, we 
smooth out (by factors of 2, to avoid splitting a grid)
the simulated data cubes of chemical abundances as well as the
synthetic observation PPV cubes from RADMC-3D to obtain $X_\CO$ at coarser
resolutions.

We impose a detection limit of
$W_\mr{CO,det}=0.75~\mr{K\cdot km/s}$ (unless specified otherwise), below which the $\CO$ emission is assumed
to be undetected. This detection limit is similar to the sensitivity
of $\CO$ observations in \citet{Sun2018}. Similar to observations, we calculate
$X_\CO$ only in the $\CO$ bright regions above the detection limit.

In addition to maps of emission in individual lines, observational studies
sometimes include two or more lines, which provide 
information regarding excitation.  
We define the ratio of the emission line intensity as
\begin{equation}
    R_{21} \equiv \frac{W_\CO(2-1)}{W_\CO(1-0)}.
\end{equation}

\section{Results}\label{section:results}
\subsection{Overall Properties\label{section:overall_properties}}

\begin{figure}[htbp]
\centering
\includegraphics[width=\linewidth]{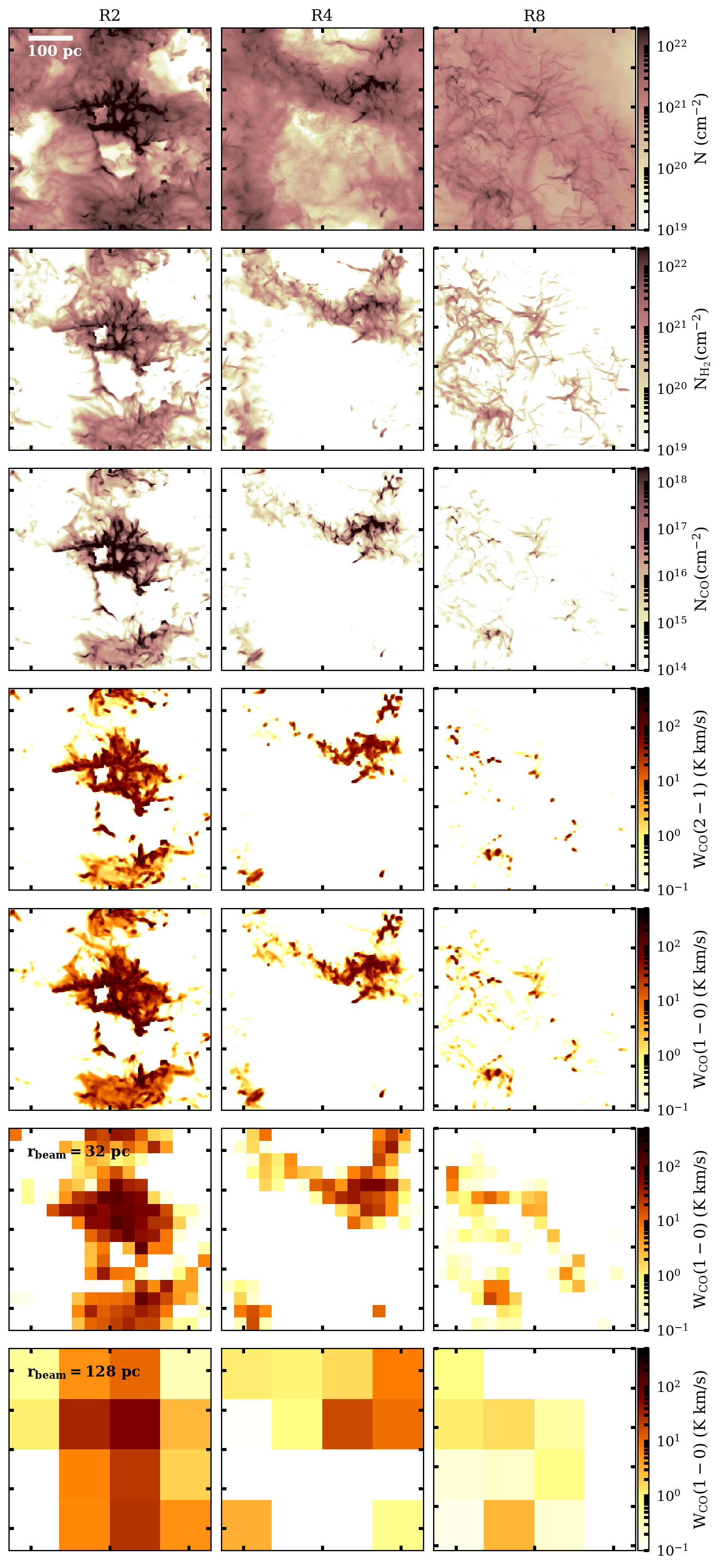}
\caption{Representative snapshots from fiducial models  R2 (R2B2-Z1, left
  column), R4 (R4-Z1, middle column), and
    R8 (R8-Z1, right column).
  Maps show: the column density of all gas ($N$, first row), 
    molecular gas ($N_\Ht$, second row), $\CO$ ($N_\CO$, third row),
    and the intensity of the CO(2--1) line ($W_\CO~(2-1)$, fourth row) and
    CO(1--0) line ($W_\CO~(1-0)$, last three rows), all viewed
    along the $z$-axis.
    The last two rows show maps $W_\CO~(1-0)$ smoothed out to larger synthetic
    beams of $r_\mr{beam}=32~\pc$ and $r_\mr{beam}=128~\pc$. All other rows
    show the maps at the original simulation resolution of 2 pc.
    The $x$ (horizontal) and $y$ (vertical) axes have a total length of 512 pc.
    The R8 model has a larger box size ($1024~\pc$), but we show a patch on the same
    scale of the R2 and R4 models for easier comparison.}
\label{fig:overview}
\end{figure}

Results from representative snapshots taken from the R2, R4 and R8 fiducial
physical models are shown in
Figure \ref{fig:overview}. As the surface density decreases from the inner
galaxy R2 model to solar neighborhood R8 model, the molecular clouds become
smaller, less dense, and fainter in $\CO$ emission.

Comparing $N_\Ht$ and
$N_\CO$, it is apparent that $\CO$ only traces the dense part of molecular
clouds. The outskirts of diffuse molecular clouds are often $\CO$-dark. This is
because $\Ht$ self-shielding of the destructive FUV radiation is 
very efficient, allowing $\Ht$ to form at lower column densities. The
formation of $\CO$, on the other hand, requires sufficient dust shielding of
the FUV radiation, which only occurs at higher column 
densities \citep{WHM2010, GOW2016}. As the surface density and density
decrease,
a larger fraction of $\Ht$ is in diffuse low density regions where $\CO$ is not
present, leading to a higher fraction of $\CO$-dark $\Ht$ (see also
Tables \ref{table:op32} and \ref{table:op128} and Section \ref{section:fdark}).

The maps of CO(2--1) and CO(1--0) line emission are very similar, with
the (2--1) line slightly fainter and tracing slightly denser gas. While 
simulations are able to produce exquisite details of turbulent molecular
clouds at $\sim\pc$ resolution, similar observational resolution is
not available in extra-galactic observations.  Even with the
unprecedented angular resolution afforded by ALMA, as in
the recent PHANGS survey, the physical resolution in galaxies beyond the
Local Group is limited to 
$\gtrsim 20~\pc$, with $\sim 100~\pc$ more typical
\citep{Leroy2016, Sun2018}. The last two rows of Figure
\ref{fig:overview} illustrate the effects of beam dilution. At 32 pc
resolution, some substructures of GMCs can still be seen. At the coarser 128 pc
resolution however, most pixels contain more than one cloud structure. The low
surface density R8 models suffer the most from beam dilution. The small and
faint clouds are smoothed out, and can fall under the observational detection
limit in some cases.

\begin{figure*}[htbp]
\centering
\includegraphics[width=0.95\linewidth]{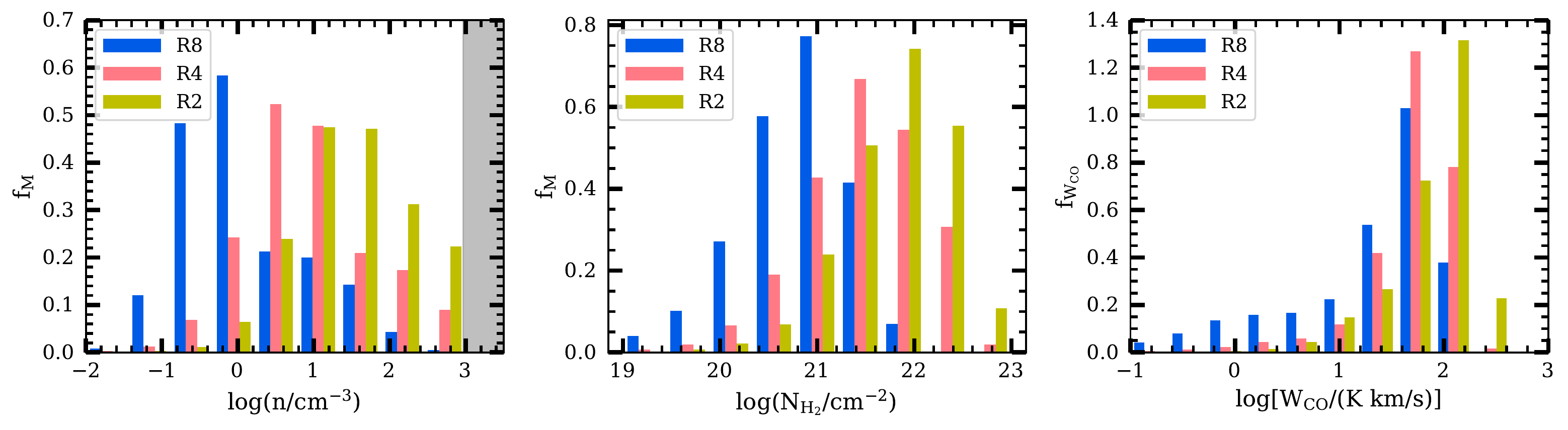}
    \caption{Mass-weighted histograms of $\log n$ (left),
    $\log N_\Ht$ (middle), and $W_\CO$-weighted histogram of
    $\log W_\CO$ ((1--0) line, right) in the snapshots shown in Figure \ref{fig:overview}
    at the original resolution of 2 pc. All histograms are normalized to have
    the same area. The gray shaded region in the left panel is
    above the critical density for sink particle creation.
    As the total surface density increases from R8 to R4 and
    to R2 models, the distributions of $n$, $N_\Ht$, and $W_\CO$ also 
    shift to higher values.  
    }
\label{fig:hist_NH2_n}
\end{figure*}

A more quantitative presentation of the gas properties for the fiducial
models is
shown in Figure \ref{fig:hist_NH2_n}. The peak of the mass-weighted density
distribution increases by about two orders of magnitude from R8 to R2 models,
and the peak of the $\Ht$ column density and CO brightness distributions
also increases by about an order of
magnitude. The higher density allows for more efficient formation of 
$\Ht$ and $\CO$ molecules and the higher surface density creates stronger
shielding of the FUV radiation field. This allows the ISM near the mid-plane to
transition from predominately atomic to predominately molecular from R8 to R2
models. We note that there is a sharp drop in the histogram of
gas density $n$ at $\gtrsim 10^3~\mr{cm^{-3}}$, which is due to the numerical
effect of sink particle creation. The peak of the density distribution, however,
is well resolved at 2 pc resolution \citepalias{Gong2018}.

A summary of the important physical and observable variables across different
models and snapshots at a synthetic beam sizes of 32 pc and 128 pc are listed in
Tables \ref{table:op32} and \ref{table:op128} in the \autoref{sec:addendix}.
Many properties of molecular
clouds vary significantly due to the changes in physical environments such as
surface density, metallicity, FUV radiation field strength, and CRIR. 
The median values of $X_{\CO,20}=0.6-3$ across different models show much less
variation than the median values of both $N_\Ht$ and $W_\CO$, showing that
$\CO$ emission traces $\Ht$ column density to some extent across all models.
However, we also note that even in a given model, there is significant
dispersion of $X_\CO$ across different regions and snapshots (as shown by the
semi-quartile ranges in brackets), sometimes up to more than 50\%.
Taken together, this variability shows the need 
to calibrate $X_\CO$ to reduce the
uncertainty in observations.

\subsection{Comparison with Observations\label{section:comparison_obs}}
\begin{figure*}[htbp]
\centering
     \begin{center}
      \subfigure{%
            \includegraphics[width=0.49\textwidth]{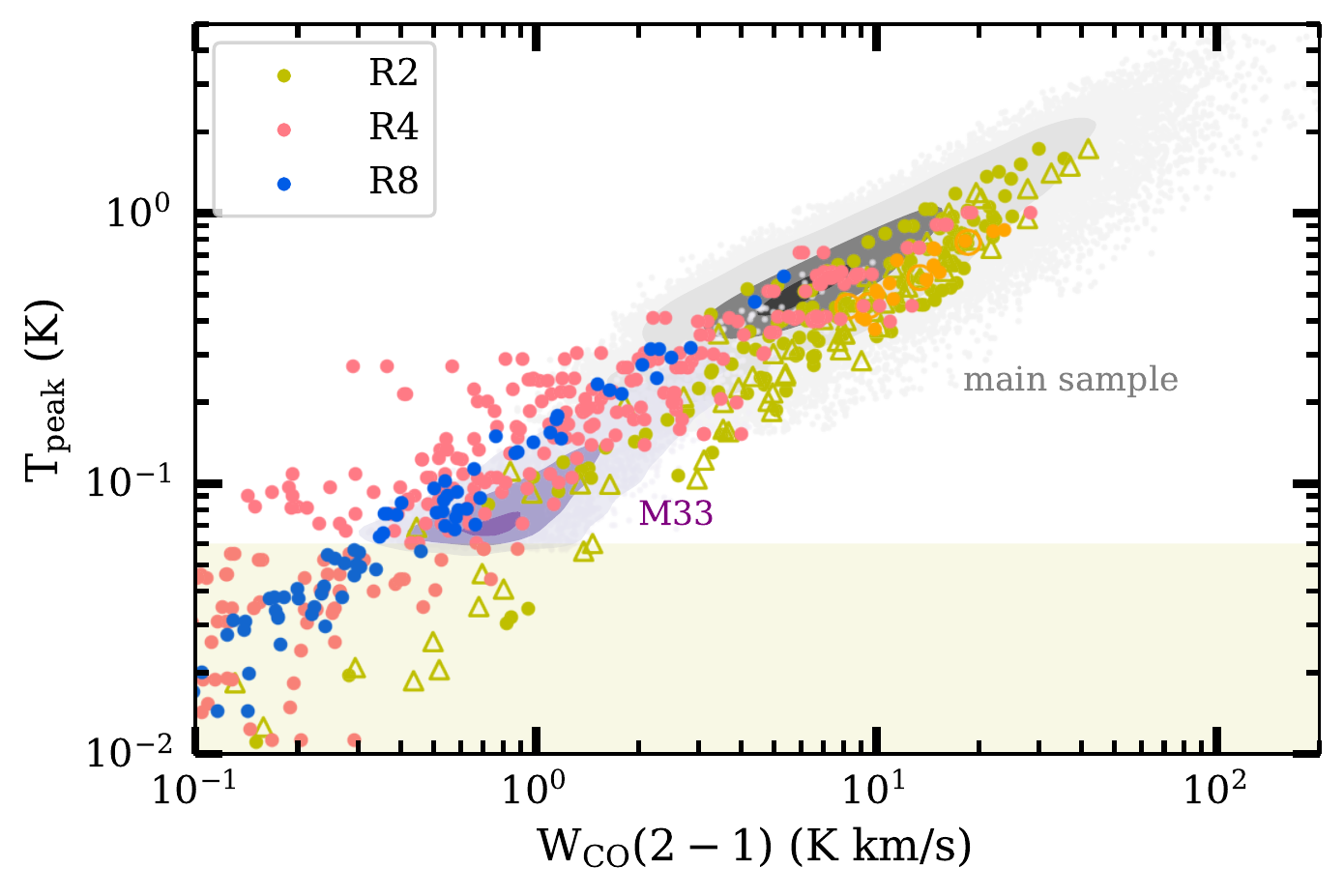}
        }
      \subfigure{%
           \includegraphics[width=0.49\textwidth]{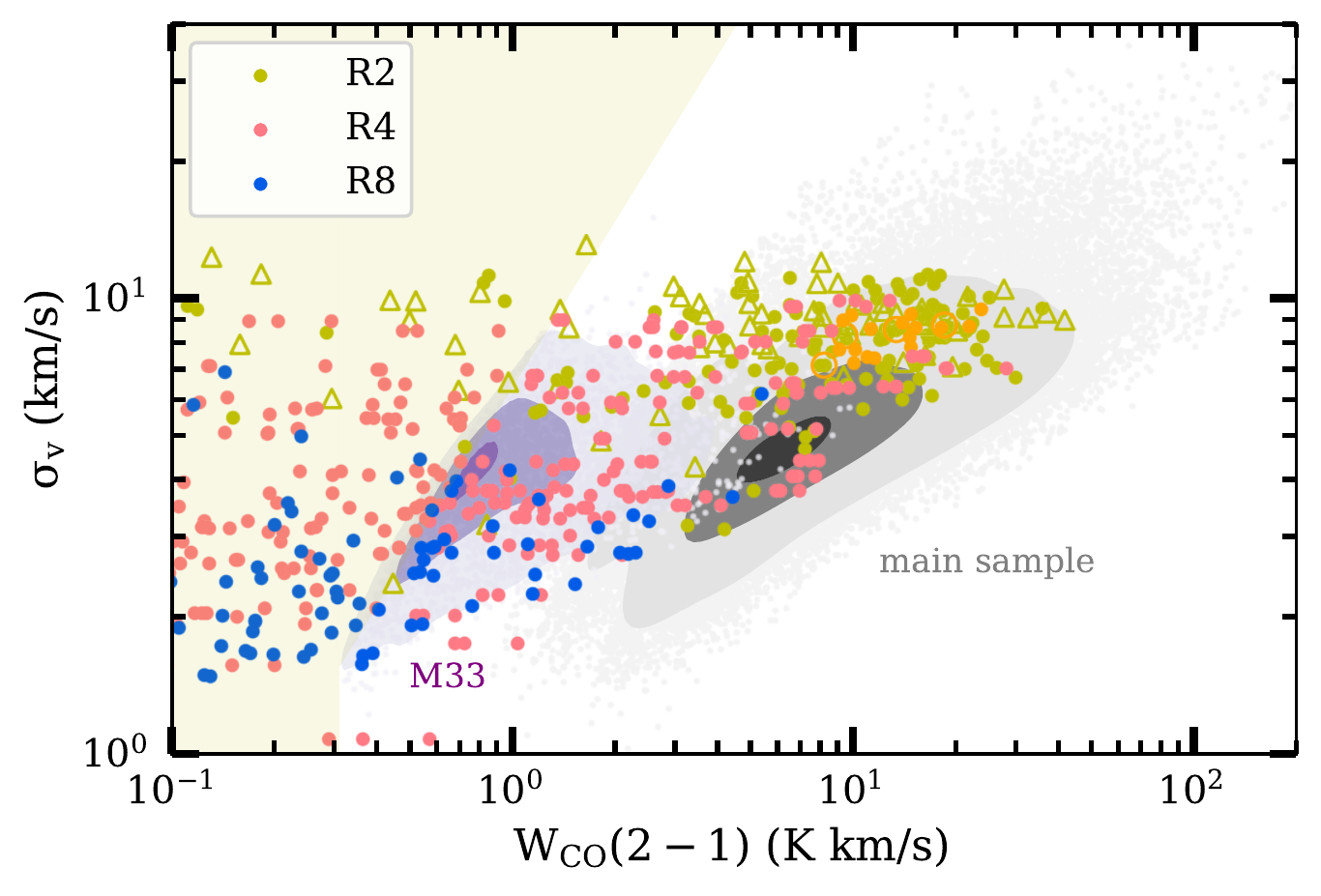}
        } 

    \end{center}
    \caption{CO (2--1) line properties at GMC scales in PHANGS observations
    and our numerical simulations. The PHANGS observations, including the main
    sample and M33 in the Local Group, are taken from
    \citet{Sun2018}, with a beam size of 120 pc. 
    Only measurements in the disk regions and above the completeness limit for
    detection are included. The contours show the PHANGS
    data density levels including 10\%, 50\%, and 90\% of the measurements. 
    The yellow shaded areas roughly mark the regions below the observational
    detection limit.
    The simulations are taken from post-processing results with solar
    metallicity $Z=1$, and a beam-size of 128 pc. 
    No detection limit is imposed in the simulations shown here, i.e, 
    the simulated data points are assumed to have a perfect sensitivity. 
    For the R2 and R4 models,
    post-processing results from different levels of FUV radiation and CRIR
    (-Z1, -Z1L10, -Z1CR10L10 models in Table \ref{table:model}) are all included,
    and their distributions are similar. For the R2 model, 
    the larger box-size model (R2B2-Z1) is shown with empty
triangles, and the higher numerical resolution model (R2N2-Z1) is shown with orange points, with the empty orange circles showing the corresponding lower
    resolution snapshot (in R2-Z1) at a similar simulation time. The
    range of physical parameters from the numerical
    simulations generally agree  with the PHANGS observations, with some points
    below the observational detection limit.
    The estimation of the data density distribution is made using the fastKDE 
    python package developed by \citet{OBrien2014, OBrien2016}.
        \label{fig:WCO_PHANGS}
       }
\end{figure*}
To validate that the molecular clouds in our simulations are realistic
representations of observed clouds, we compare our simulation results to
the cloud properties directly obtained from $\CO$ observations, such as
$W_\CO$, $\sigma_v$, $T_\mr{peak}$ and $R_{21}$.

Figure \ref{fig:WCO_PHANGS}
compares the molecular cloud properties traced by the CO(2--1) emission observed in the PHANGS galaxies
(120 pc beam) with those in the synthetic observations from our simulations (128 pc beam).
The simulations successfully reproduce both the correlations between and the
range of the observed $W_\CO$, $T_\mr{peak}$ and $\sigma_v$. This confirms that
the molecular clouds in our simulations are indeed realistic. Because we only
simulate patches of galaxies, and do not account for the whole galactic
environment, we cannot match the detailed statistical distribution of the
observables in PHANGS. Our simulations suggest that many molecular clouds
exist below the detection limit of PHANGS, especially in the lower surface
density environments represented by the R4 and R8 models. The differences in the
cloud properties observed in the nearby M33 and the main sample of PHANGS 
are at least partly due to the limited observational sensitivity. Even in our
highest surface density model R2, many fainter clouds exist below the
detection limit of the main sample in PHANGS, and the distribution smoothly
extends to those observed in M33.

\begin{figure*}[htbp]
\centering
\includegraphics[width=\linewidth]{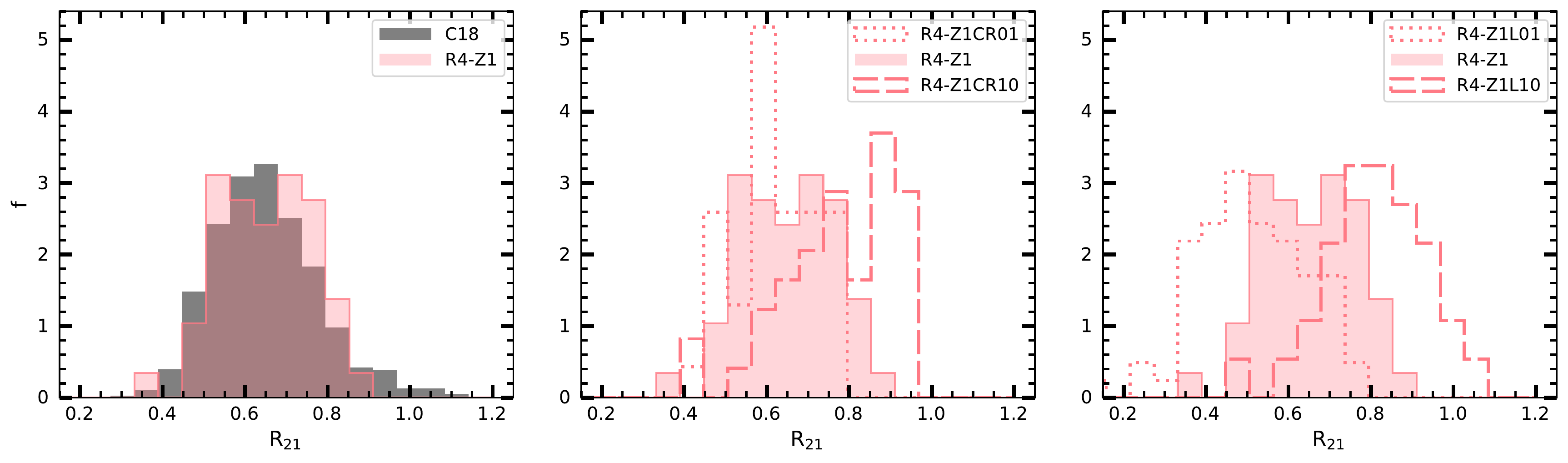}
\caption{
Line ratio $R_{21} = W_\CO(2-1)/W_\CO(1-0)$ in comparison to observations, and for varying radiation and cosmic ray  conditions. {\sl Left panel:} Normalized histogram of $R_{21}$ for fiducial R4 model (R4-Z1) in comparison to the observations by \citet{Cormier2018} (C18).  
The observations of C18
have a spatial resolution of $\sim 1.5 ~\mr{kpc}$ and we show the simulated
    $R_{21}$ averaged over the whole 512 pc box, with the histogram showing
    the distribution from all temporal snapshots.  The model
    reproduces the wide range of $R_{21}$ observed in C18.
    {\sl Middle and right panel:} $R_{21}$ variations associated with
    variations in CRIR and FUV
    radiation. The middle panel shows the normalized histograms in models with CRIR 
    10 times lower (R4-Z1CR01) and higher (R4-Z1CR10)  compared to the
    fiducial model R4-Z1. The right panel shows models with  incident FUV radiation 
    10 times lower (R4-Z1L01) and higher (R4-Z1L10).
    Increasing either form of radiation moves the peak of the distribution
    to higher $R_{21}$.
    }
\label{fig:C18_R4}
\end{figure*}

Figure \ref{fig:C18_R4} shows the comparison of $R_{21}$ between our
simulations and nearby spiral galaxies observed in the EMPIRE survey
\citep{Cormier2018}. Most regions covered by the EMPIRE survey have a total 
gas surface density of $15-50~M_\odot\mr{pc^{-2}}$ and star formation rate of 
$0.01-0.1~M_\odot\mr{yr^{-1}kpc^{-2}}$ \citep{Cormier2018, Jimenez-Donaire2019}. 
This is closest to the R4 environment in our simulations,
and thus we plot the R4-Z1 model for comparison. The left panel of
Figure \ref{fig:C18_R4} shows that we successfully reproduce the observed
distribution of $R_{21}$. This is a significant improvement over the one-zone
model RADEX \citep{RADEX2007}, which fails to reproduce the wide
range of $R_{21}$ observed \citep[see Figure 7 in][]{Cormier2018}. The middle
and right panels of Figure \ref{fig:C18_R4} illustrate that increasing either the
FUV radiation strength or
CRIR tends to increase $R_{21}$. Qualitatively, this can
be understood because both FUV radiation and CRs preferentially destroy CO in lower density gas,
causing most of the $\CO$ emission to occur at higher densities, where $R_{21}$
is also higher on average. The mean values of $R_{21}$ in R4-Z1L01, R4-Z1 and 
R4-Z1L10, for which the background radiation field increases from 0.1 to 1 to
10 times the fiducial value, are 0.51, 0.65 and 0.82. A simple linear fit
between $\log (J_\mr{FUV})$ and $R_{21}$ gives a slope of 0.152, close to the
slope of 0.161 found in observations of M83 by \citet{Koda2020}.

\subsection{$X_\CO$ conversion factor\label{section:results_XCO}}
\subsubsection{Dependence on
  Metallicity, FUV Radiation and Cosmic Rays}\label{section:XCO_env}
\begin{figure*}[htbp]
\centering
\includegraphics[width=\linewidth]{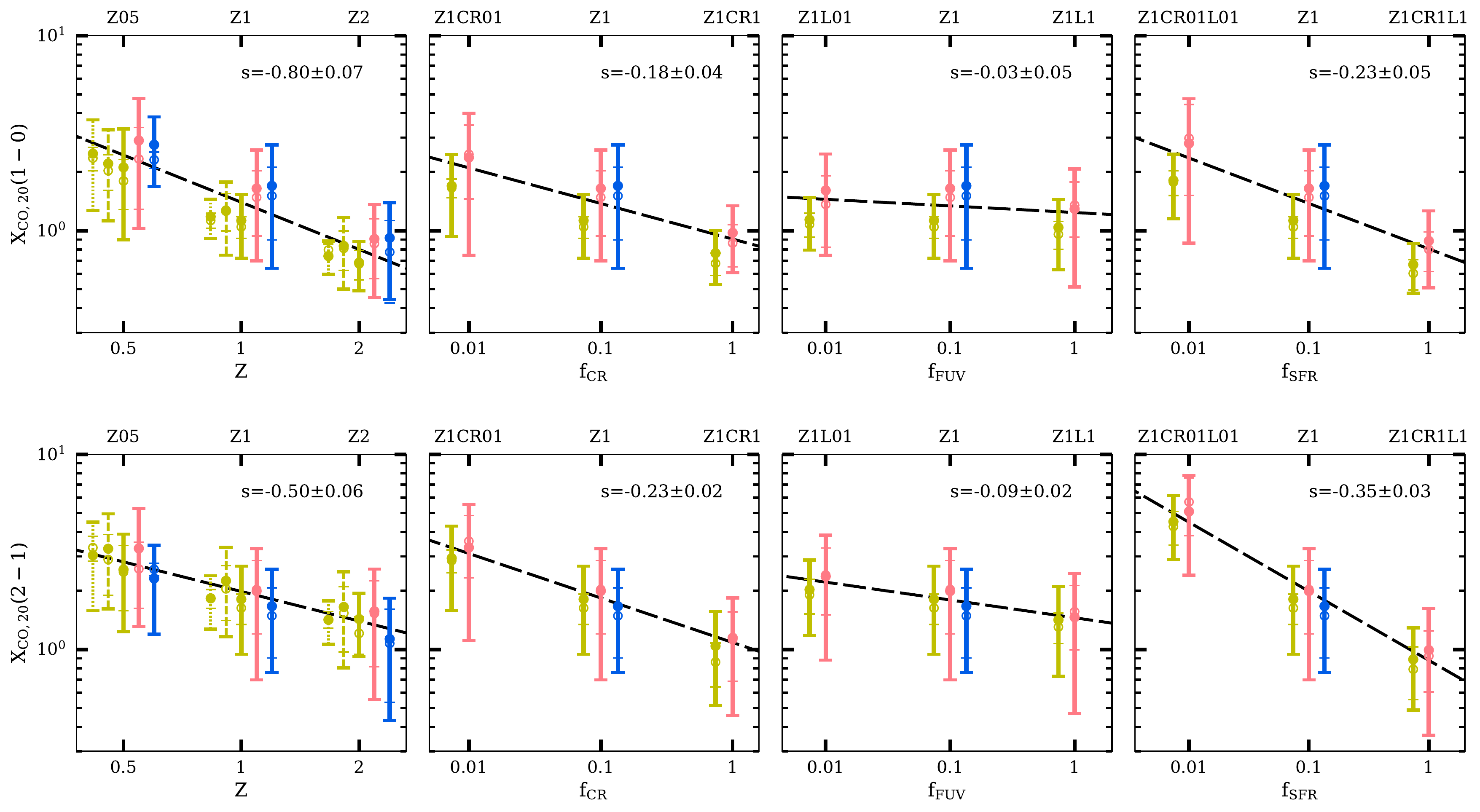}
\caption{The $X_\CO$ conversion factor for CO(1--0) (upper panels) and
  CO(2--1) (lower panels) lines. The x-axes correspond to parameter values encoded  in  model IDs, as given in Table \ref{table:model}; points in each
  group are slightly  offset to left and right  for clarity.
  Results for
  the R2, R4, and R8 models are shown in yellow,
  red, and blue colors, respectively. Points to the left of the main R2
  simulations are from R2B2 (larger box size) and R2N2 (higher resolution)
  models, shown with dashed and dotted line styles.
For all models, symbols and error bars show the median value
$X_\CO$ and the semi-quartile range of $X_\CO$ in $\CO$-bright regions with
a 32 pc (filled circle, thick error bar) and 128 pc (empty circle, thin
error bar) beam (see also Tables \ref{table:op32} and \ref{table:op128}). 
For each panel, the black dashed line shows a linear fit of $\log X_\CO$ (median values at
32 pc resolution, shown as the filled circles) as a function of the
environmental parameters $\log Z$, $\log \xi_0$, $\log \chi_0$ and 
$\log f_\mathrm{SFR}$ (all models shown in each panel are included in the fits). Fitting with median values at 128 pc resolution gives very similar slopes.
The fitted value of the slope and its standard
deviation is written in the corresponding panel.
Evidently, the main environmental drivers for the variation in $X_\CO$ are metallicity and
the CRIR.
\label{fig:XCO}
}
\end{figure*}

Figure \ref{fig:XCO}  summarizes results of $X_\CO$ from all of our
models for both the $J=1-0$ (top) $J=2-1$ (bottom) lines,
separately showing variations due to
metallicity, FUV radiation, and CRIR when the parameters $Z$,
$f_\mathrm{CR}$, $f_\mathrm{FUV}$ are independently
varied, and when the last two are varied together as $f_\mathrm{SFR}$. The
R2B2 (larger box size) and R2N2 (higher resolution) models
have very similar $X_\CO$ to the fiducial R2 model. This
confirms that $X_\CO$ is converged at the current box size and 2 pc resolution, 
as previously found in \citetalias{Gong2018}.

\autoref{fig:XCO} also shows results of fitting the variation of $X_\CO$ with
varying metallicity ($Z$), CRIR ($\xi_0$), and background FUV strength ($\chi_0$).  
As expected (see Section \ref{section:theory}), $X_\CO$ decreases with
increasing $Z$. 
It is interesting that the measured scalings $X_\CO (1-0)\propto Z^{-0.8}$ and
$X_\CO (2-1)\propto Z^{-0.5}$ are similar to the  relation $X_\CO \propto Z^{-1/2}$
predicted based on a highly simplified
model in \autoref{eq:XCO_approx1}, under the assumption $n_\CO \propto Z$ in
CO-emitting  regions.
The physical reason for the increase of $X_\CO$ at lower $Z$ is the decreased excitation temperature due to lower optical depth of CO lines (see also Figure \ref{fig:Tgas_Texc_CRL} and related text), although the lines are still optically thick. Compared to the (1--0) line, the (2--1) line traces denser gas where the CO abundance is less sensitive to the change in dust shielding (as the shielding is already above the critical values required for CO formation), and thus shows a weaker dependence on $Z$.

Also consistent with general expectations,
considering the  decrease
of $X_\CO$ at higher  $T_\mr{gas}$ (see \autoref{eq:XCO_virial} and \autoref{eq:Texc}) and the increase of $T_\mr{gas}$ at higher CRIR in shielded regions,
$X_\CO$ decreases roughly $\propto \xi_0^{-0.2}$.  There  is also very weak
dependence on
the FUV radiation field,  roughly $X_\CO(1-0) \propto \chi_0^{-0.03}$ or $X_\CO(2-1) \propto \chi_0^{-0.09}$.
This insensitivity is reasonable, given that FUV mainly
affects the gas volume and mass where CO and $\Ht$ can form (limited by
photodissociation), rather than the conditions in shielded regions.  

Since $Z$ is often readily available in observations, the  fits shown in 
\autoref{fig:XCO}  (see also 1a and 1b in \autoref{table:XCO_fit})
can be used to calibrate $X_\CO$ in different galactic environments. While
the dependence of $X_\CO$ on the CRIR is also quite clear from our simulations,
the value $\xi_0$ is not easily accessible observationally.  Since the
physical dependence on $\xi_0$
is expected to be mainly through the gas temperature, which
affects excitation, other avenues to controlling for this effect are available.
We discuss this further below.

\begin{figure*}[htbp]
\centering
\includegraphics[width=\linewidth]{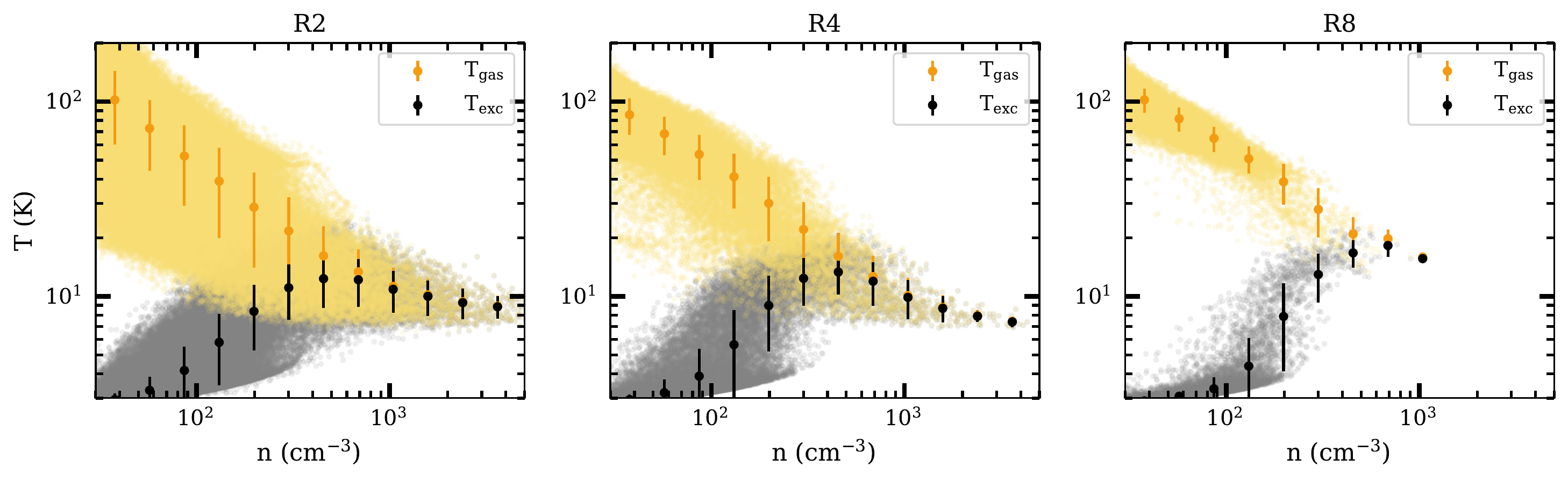}
    \caption{Gas temperature $T_\mr{gas}$ (orange) and CO(1--0) line excitation
    temperature $T_\mr{exc}$ (black) from the snapshots shown in Figure
    \ref{fig:overview}. The binned average values with standard deviations are
    shown together with the background scatter of individual pointings.
    The excitation temperature
    approaches the gas temperature at densities $n \gtrsim 500~\mr{cm^{-3}}$.
    The $T_\mr{exc}$ -- $n$ relation is important for $X_\CO$ 
    (\autoref{eq:XCO_theory} and \ref{eq:XCO_virial}).}
\label{fig:Tgas_Texc}
\end{figure*}

\begin{figure}[htbp]
\centering
\includegraphics[width=\linewidth]{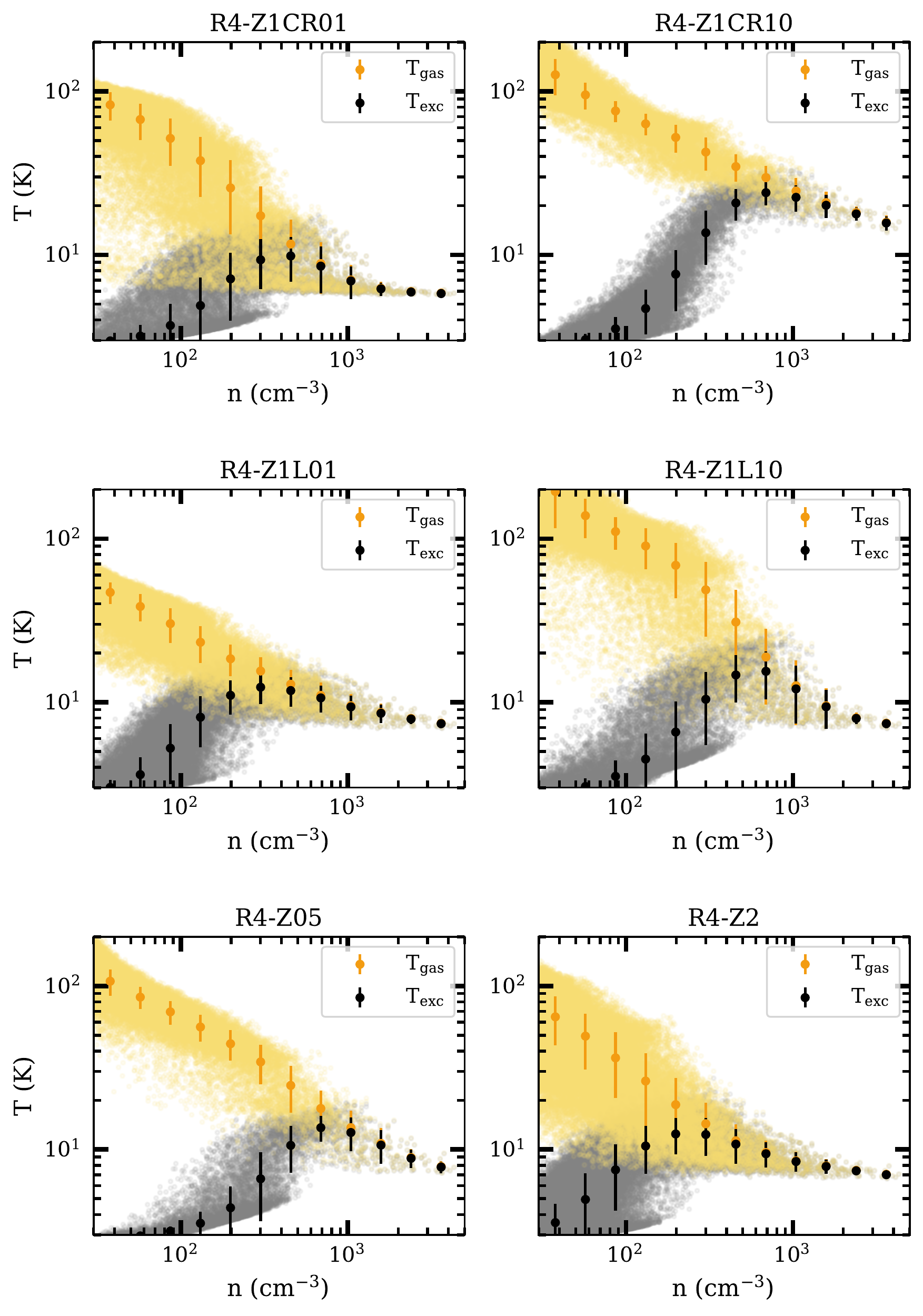}
\caption{Dependence on density of $T_\mr{gas}$  and $T_\mr{exc}$ as in 
  \autoref{fig:Tgas_Texc}, but for just model R4 at varying CRIR,
   FUV intensity, and metallicity. Compared
  to the middle panel of \autoref{fig:Tgas_Texc}, here we  show: 
  (top row)  CRIR 10 times lower (R4-Z1CR01) or higher (R4-Z1CR10); (middle row)   incident FUV
    radiation
10 times lower (R4-Z1L01) or higher (R4-Z1L10); (bottom row) metallicity 2 times lower (R4-Z05) and higher  (R4-Z2).}
\label{fig:Tgas_Texc_CRL}
\end{figure}

Motivated by the theoretical expectations
(see Equations \ref{eq:XCO_virial}, \ref{eq:Texc}, and \ref{eq:beta_nc}),
we further examine the relation between $T_\mr{exc}$
and $n$ in Figures \ref{fig:Tgas_Texc} and \ref{fig:Tgas_Texc_CRL}. At low
density, there is a large difference between $T_\mr{exc}$ and $T_\mr{gas}$.
As the density increases, $T_\mr{exc}$ increases both due to the higher
collisional rates and the increased optical depth. At the same time,
$T_\mr{gas}$ decreases due to decreased heating from the shielding of the FUV
radiation  (and CRs),
and increased cooling at higher densities. As pointed out by 
\citet{GOW2016}, because FUV radiation dissociates $\CO$, the $\CO$-rich
regions are generally shielded by high columns of dust,
and CR ionization dominates heating of the gas. At high enough density
(cf. Equations \ref{eq:Texc} and \ref{eq:XCO_approx2}), 
$T_\mr{exc}$ reaches LTE with $T_\mr{gas}$. In  shielded gas,
$T_\mr{gas}$ is mostly set by the CRIR,
and decreases slightly at high densities due to the decrease in low-energy
cosmic  rays penetrating to high columns (following our adopted relation in 
\autoref{eq:CRatten}). Although
$\xi_0$ is higher in R2 models, there is also more shielding due to
the higher surface density (see also Table \ref{table:model}).
As a result, the CRIR and temperature in the $\CO$ 
dominated gas are similar across the fiducial R2, R4 and R8 models. 
At lower densities where $T_\mr{exc} < T_\mr{gas}$,
the $T_\mr{exc}$ values in R2 models are slightly higher due to the higher
optical depth. This leads to the slightly lower $X_\CO$ in the fiducial R2
models (Equation \ref{eq:XCO_virial}).

Figure \ref{fig:Tgas_Texc_CRL} further examines the $T_\mr{exc}$ -- $n$ relation in
models with varying CRIR, FUV radiation and metallicity. Increasing the
CRIR (top row) leads to higher temperature in the dense, shielded regions, resulting in 
higher $T_\mr{exc}$; this is the reason for the decrease of $X_\CO$ at higher
$f_\mr{CR}$ seen in \autoref{fig:XCO}. An increase in the  FUV radiation
(second row) also increases the gas temperature, but only in the low density
and minimally shielded gas.
At the same time,  photodissociation of $\CO$ decreases the optical
depth. These two effects tend to cancel each other, and as a result, the
$X_\CO$ is relatively insensitive to the FUV radiation
(as seen in the weak dependence on $f_\mr{FUV}$ in \autoref{fig:XCO}).
Increasing metallicity (third row)
leads to more shielding and more efficient $\CO$ formation.  At low (high) $Z$,
line saturation -- with $T_\mr{exc}$ approaching $T_\mr{gas}$ --
occurs at higher (lower) densities. Overall, an increase in $Z$ results in
higher optical depth, higher $T_\mr{exc}$, and lower $X_\CO$. 

\begin{figure}[htbp]
\centering
\includegraphics[width=\linewidth]{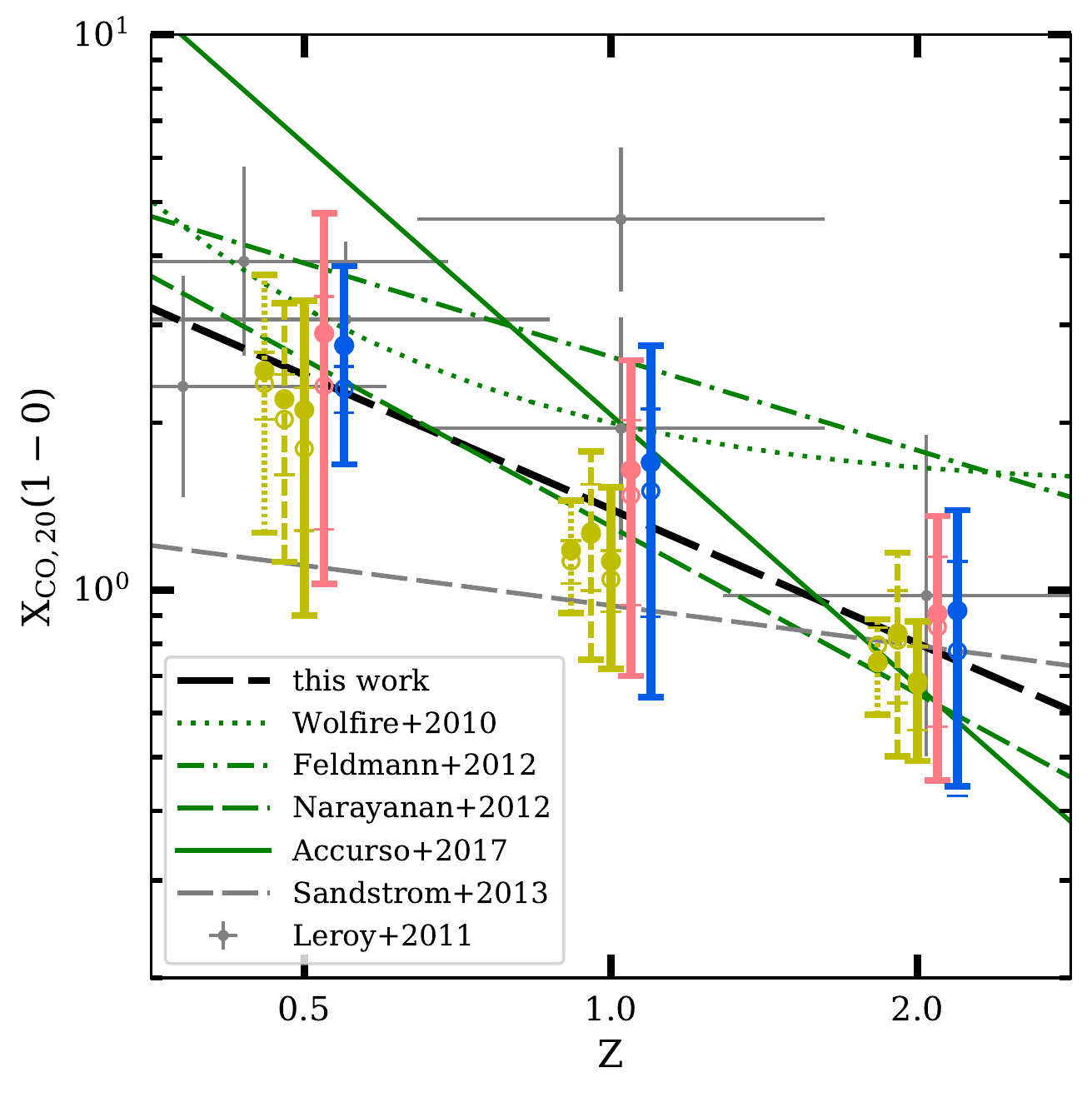}
    \caption{Summary of results for $X_\CO(1-0)$ versus $Z$.
    The yellow, red and blue error bars and the black dashed line show the
    results from our fiducial R2, R4 and R8 models as in
    the top left panel of Figure \ref{fig:XCO}, with slight horizontal
    offsets to avoid overlaps.   We compare to 
    other theoretical predictions (green lines) and observations
    (gray lines and symbols), as follows.
    \citet{WHM2010}: PDR models.
    \citet{Narayanan2012} and \citet{Feldmann2012}:
    Galaxy simulations with sub-grid models for
    molecular clouds. 
    \citet{Accurso2017}: numerical models of 
    spherically symmetric star forming regions.
    \citet{Leroy2011}: observations of local group galaxies,
    averaged over large areas comparable to the size of the galaxy; 
    $\Ht$ mass
    from dust. \citet{Sandstrom2013}: nearby spiral and
    dwarf galaxies, averaged over kilo-parsec scale; $\Ht$ mass from dust.
    Our results and fit $X_\CO(1-0) \propto Z^{-0.8}$ in the range
    of $Z=0.5-2$ are consistent with other theoretical predictions and
    observations.
    }
\label{fig:XCO_Z_obs}
\end{figure}

Of the ``environmental'' factors affecting $X_\CO$, the dependence on $Z$
has been the most extensively studied in theory and observations.   
We show a comparison between our results and recent
literature in Figure \ref{fig:XCO_Z_obs}. Among the theoretical studies
shown, our work is the only one that has resolved
clouds forming (and dispersing)
in time-dependent simulations of the multiphase ISM
with self-consistent star formation and feedback.
The slope of $-0.8$ found by us for  $X_\CO(1-0)$ lies
in between other theoretical predictions. Our values of $X_\CO$ are also
consistent with observations of the Milky Way and nearby galaxies. We note that
our results are only valid between $Z=0.5-2$. The MHD simulations are
run with $Z=1$, and a large departure from $Z=1$ can change the dynamical
structure of the clouds where molecules form
by changing the efficiency of heating and
cooling. Furthermore, at lower metallicities, decreased shielding causes $\CO$
to form at higher densities, which would  require
higher numerical resolution. We have 
experimented with setting $Z=0.1$, and found that current resolution of 1 --
2 pc is inadequate in order to resolve $X_\CO$.

\subsubsection{Dependence on Physical Properties of the 
Gas\label{section:XCO_physical}}
While in Section \ref{section:XCO_env} we investigate
the variation of average $X_\CO$ on large scales associated with key
environmental factors, in this
section we consider the variation of $X_\CO$ on small scales due to
the structure and spatially-varying conditions
within  molecular clouds. 

\begin{figure}[htbp]
\centering
\includegraphics[width=\linewidth]{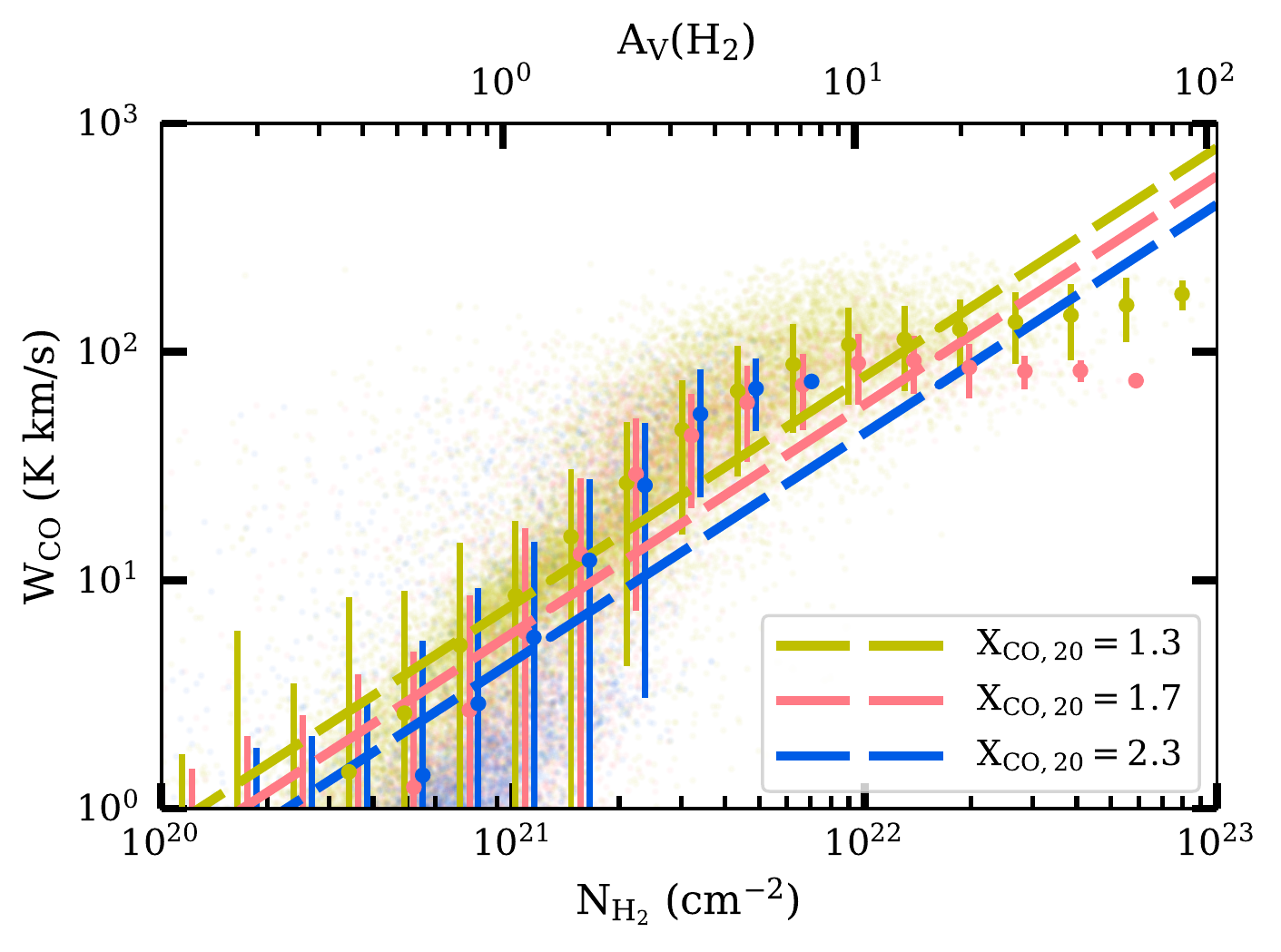}
    \caption{$N_\Ht$ versus $W_\CO(1-0)$ for the R2-Z1 (yellow), R4-Z1 (red) 
    and R8-Z1 (blue) snapshots shown in Figure \ref{fig:overview} at the native
    simulation resolution of 2 pc. The binned mean values and
    standard deviations are
    plotted over the background of scattered individual points. The
    dashed lines show the average $X_\CO$ in the CO-bright 
    ($W_\CO > 0.75~\mr{K\cdot km/s}$) regions for each model.
    }
\label{fig:NH2_WCO}
\end{figure}

\begin{figure*}[htbp]
\centering
\includegraphics[width=\linewidth]{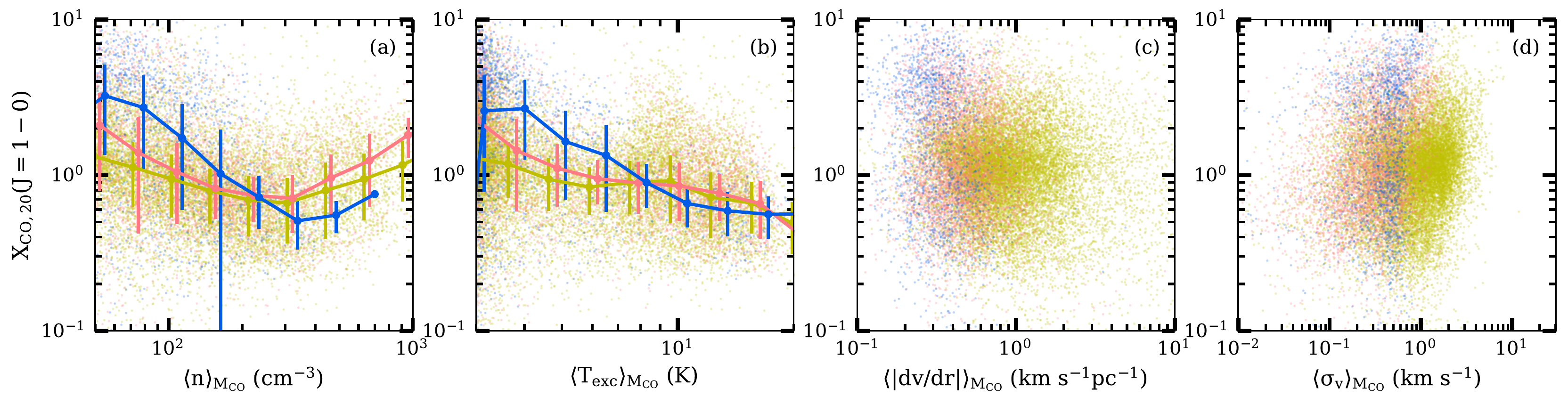}
    \caption{Correlation between $X_\CO(1-0)$ and physical properties of the
    gas for the R2-Z1 (yellow), R4-Z1 (red) and R8-Z1 (blue) snapshots shown in
    Figure \ref{fig:overview}. The parameters
    $\langle n \rangle_{M_{CO}}$,
    $\langle T_\mr{exc} \rangle_{M_{CO}}$, $\langle |\di v/\di r| \rangle_{M_{CO}}$, and 
    $\langle \sigma_{v} \rangle_{M_{CO}}$
    are the gas density, excitation temperature of the $J=1-0$ transition, 
    the velocity gradient, and velocity dispersion along the line of sight,
    weighted by the $\CO$ mass. Each point represents a pixel at the native
    simulation resolution of 2 pc. The binned median values and semi-quartile
    range are plotted over the background of scatter points for the left two
    panels (medians are not shown in the right two panels, where
    no significant correlation is found).  Only
    pixels with $W_\CO > 2~\mr{K\cdot km/s}$ are shown.
    }
\label{fig:XCO_correlation_physical}
\end{figure*}

\begin{figure}[htbp]
\centering
\includegraphics[width=\linewidth]{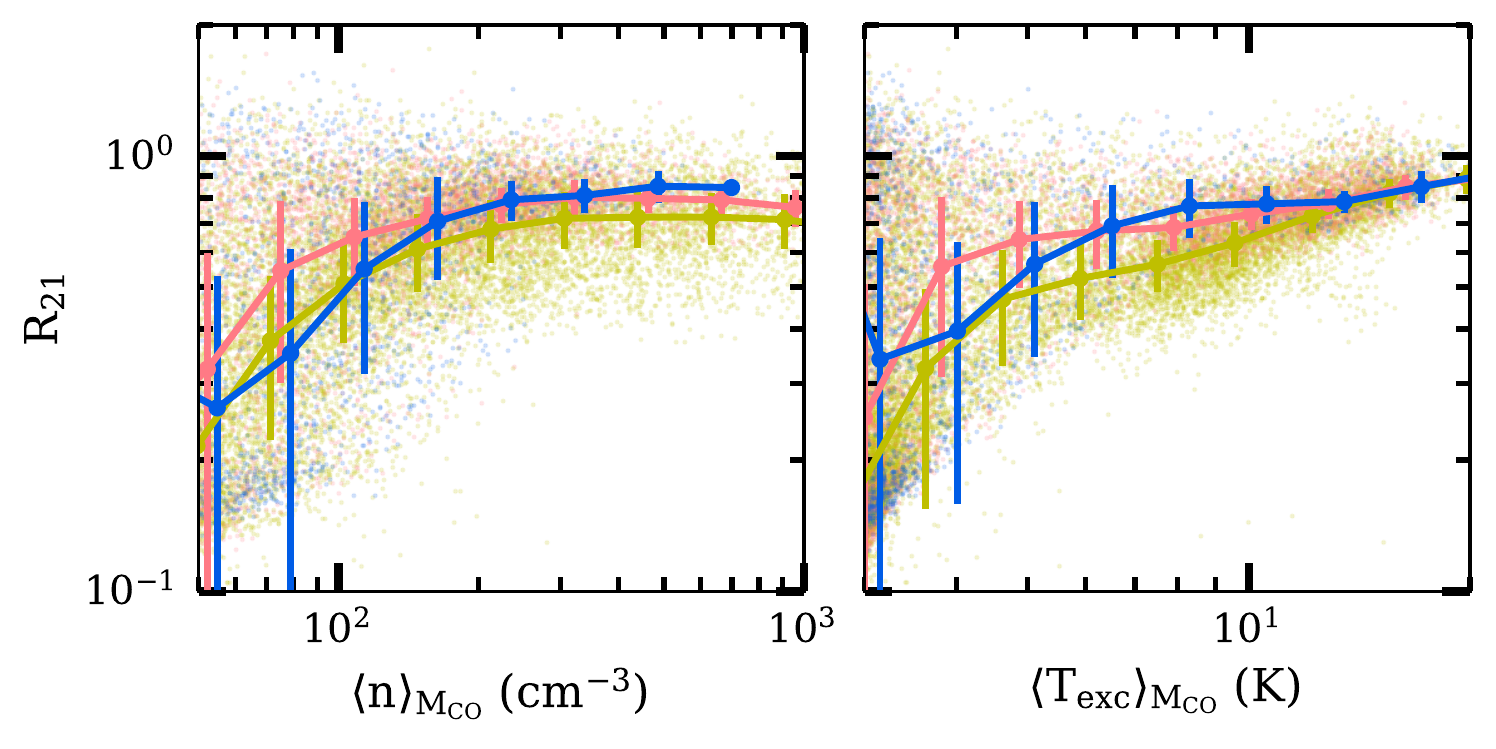}
    \caption{Correlation between $R_{21}$ and physical properties of the
    gas, similar to Figure \ref{fig:XCO_correlation_physical}. 
    }
\label{fig:R21_correlation_physical}
\end{figure}

First, it is evident from the $W_\CO$ -- $N_\Ht$ relation illustrated in
\autoref{fig:NH2_WCO}
that $X_\CO$ systematically varies with
surface density at small scales within
molecular clouds. On the one hand, at low $N_\Ht$ the $\CO$ abundance is low
due to photodissociation at low $A_V$, whereas $\Ht$ is non-negligible, being
self-shielded.  On the other hand, at high $N_\Ht \gtrsim 5\times 10^{21}~\mr{cm^{-2}}$
the relation  flattens as
$W_\CO$ saturates due to the high optical depth. As a result, 
the resolved $W_\CO$ vs. $N_\Ht$ relations are steeper than the
large-scale averages (shown as dashed lines)
in the range $N_\Ht \sim 0.7- 5\times 10^{21}~\mr{cm^{-2}}$. 
To obtain the correct $N_\Ht$, an $X_\CO$ higher than the large-scale
average would be required at $N_\Ht \lesssim 2\times 10^{21}~\mr{cm^{-2}}$
($A_V \lesssim 2$), whereas an $X_\CO$ lower than the large-scale
average would be required at  
 $N_\Ht \gtrsim 2 \times 10^{21}~\mr{cm^{-2}}$ ($A_V \gtrsim 2$). 
Similar trends are also found in high resolution observations of
local molecular clouds \citep{Pineda2008, Lee2018}, simulations of
individual molecular clouds \citep{Shetty2011,Shetty2011b,Szucs2016}
and zoom-in simulations \citep{Seifried2020}.

Inspired by Equations \ref{eq:XCO_theory} -- \ref{eq:XCO_approx2}, we
investigate the correlation between $X_\CO$ and physical properties of the gas
on small scales in Figure \ref{fig:XCO_correlation_physical}.
The left panel
directly shows that $X_\CO$ first decreases and then increases with density,
consistent with the theoretical expectations from
Equations \ref{eq:XCO_approx1} and \ref{eq:XCO_approx2}. The $X_\CO$ --
$T_\mr{exc}$ relation shown in the second panel
can be explained by reference to \autoref{eq:Texc_approx} and
\autoref{eq:XCO_approx1}.
If $f_\CO$ and $|\di v/\di r|$ are constant or have no systematic variation
in $\CO$-bright regions, then $T_\mr{exc}\propto n$ and 
$X_\CO \propto {T_\mr{exc}}^{-1/2}$. The right two panels  of
\autoref{fig:XCO_correlation_physical} show that $X_\CO$ is 
uncorrelated with the local velocity gradient $|\di v/\di r|$
and the large scale velocity dispersion along the line of sight. 

\autoref{fig:R21_correlation_physical}
examines the relation between $R_{21}$ and gas properties.
$R_{21}$ is high at higher $n$
and $T_\mr{exc}$, and has a large scatter at lower $n$ and $T_\mr{exc}$. This
is consistent with the observations by \citet{Koda2020}, who found that
$R_{21}$ has a large spread in regions with low $W_\CO$, and $R_{21}$ 
is high in regions with high $W_\CO$.
Because $R_{21}$ correlates with $n$ and $T_\mr{exc}$, it also
correlates with $X_\CO$, and we use this to calibrate $X_\CO$ in Section
\ref{section:calibration}.

\subsubsection{Calibrating $X_\CO$ Using Observable  Quantities}
\label{section:calibration}

\begingroup
\renewcommand{\arraystretch}{2.0}
\begin{table*}[htbp]
    \caption{Fitting Results: $X_\CO$ as a function of observables\tablenotemark{a}}
    \label{table:XCO_fit}
    \centering
    \begin{tabular}{l cc c}
        \tableline
        \tableline
        Number &Transition &Parameters &Fitting Result \\
        \tableline
        1a &1-0 &$Z$ 
        &$X_\mr{CO,20}=1.4Z^{-0.80}$ \\
        1b &2-1 &$Z$ 
        &$X_\mr{CO,20}=2.0Z^{-0.50}$ \\
        2a &1-0 &$R_{21}, Z, r_\mr{beam}$ 
        &$X_\mr{CO,20}=0.93(R_{21}/0.6)^{-0.87}Z^{-0.80}
        (\mr{min}\{r_\mr{beam},100\})^{0.081}$ \\
        2b &2-1 &$R_{21}, Z, r_\mr{beam}$ 
        &$X_\mr{CO,20}=1.5(R_{21}/0.6)^{-1.69}Z^{-0.50}
        (\mr{min}\{r_\mr{beam}, 100\})^{0.063}$ \\
        3a &1-0 &$T_\mr{peak}, Z, r_\mr{beam}$ 
             &$X_\mr{CO,20}=1.8T_\mr{peak}^{-0.64+0.24\log r_\mr{beam}}
             Z^{-0.80}r_\mr{beam}^{-0.083}$ \\ 
        3b &2-1 &$T_\mr{peak}, Z, r_\mr{beam}$ 
             &$X_\mr{CO,20}=2.7T_\mr{peak}^{-1.07+0.37\log r_\mr{beam}}
             Z^{-0.50}r_\mr{beam}^{-0.13}$\\ 
        4a &1-0 &$W_\CO, Z, r_\mr{beam}$ &$X_\mr{CO,20}=6.1 
            W_\CO^{-0.54+0.19\log r_\mr{beam}}
            Z^{-0.80}r_\mr{beam}^{-0.25}$\\
        4b &2-1 &$W_\CO, Z, r_\mr{beam}$ &$X_\mr{CO,20}=21.1 
            W_\CO^{-0.97+0.34\log r_\mr{beam}}
            Z^{-0.50}r_\mr{beam}^{-0.41}$\\
        \tableline
        \tableline
    \end{tabular}
    \tablenotetext{1}{The fits are performed using the least-squares method and
    using data in $\CO$-bright regions
    from the synthetic observations in models R[2,4,8]-Z[05,1,2] and
    R2B2-Z[05,1,2]. Expressions 1a/b are from fitting the median values of
    $X_\CO$ in Figure \ref{fig:XCO}. The rest are from fitting individual
    pixels at $r_\mr{beam} = 2-128~\mr{pc}$ and with fixed slopes for $Z$ dependence from expressions 1a/b. 
    The fits are applicable to the range of $W_\CO = 0.75-200~\mr{K\cdot km/s}$.  
    The units of the physical variables are as
    follows: $W_\CO$ in $\mr{K\cdot km/s}$, $T_\mr{peak}$ in $\mr{K}$, and
    $r_\mr{beam}$ in $\mr{pc}$. For $r_\mr{beam}\gtrsim 100~\mr{pc}$, $X_\CO$ does not correlate with $W_\CO$ or $T_\mr{peak}$ due to beam dilution,
    and the beam-size independent expressions 1a/b or 2a/b should be used.
    }
\end{table*}
\endgroup

\begin{figure*}[htbp]
\centering
\includegraphics[width=\linewidth]{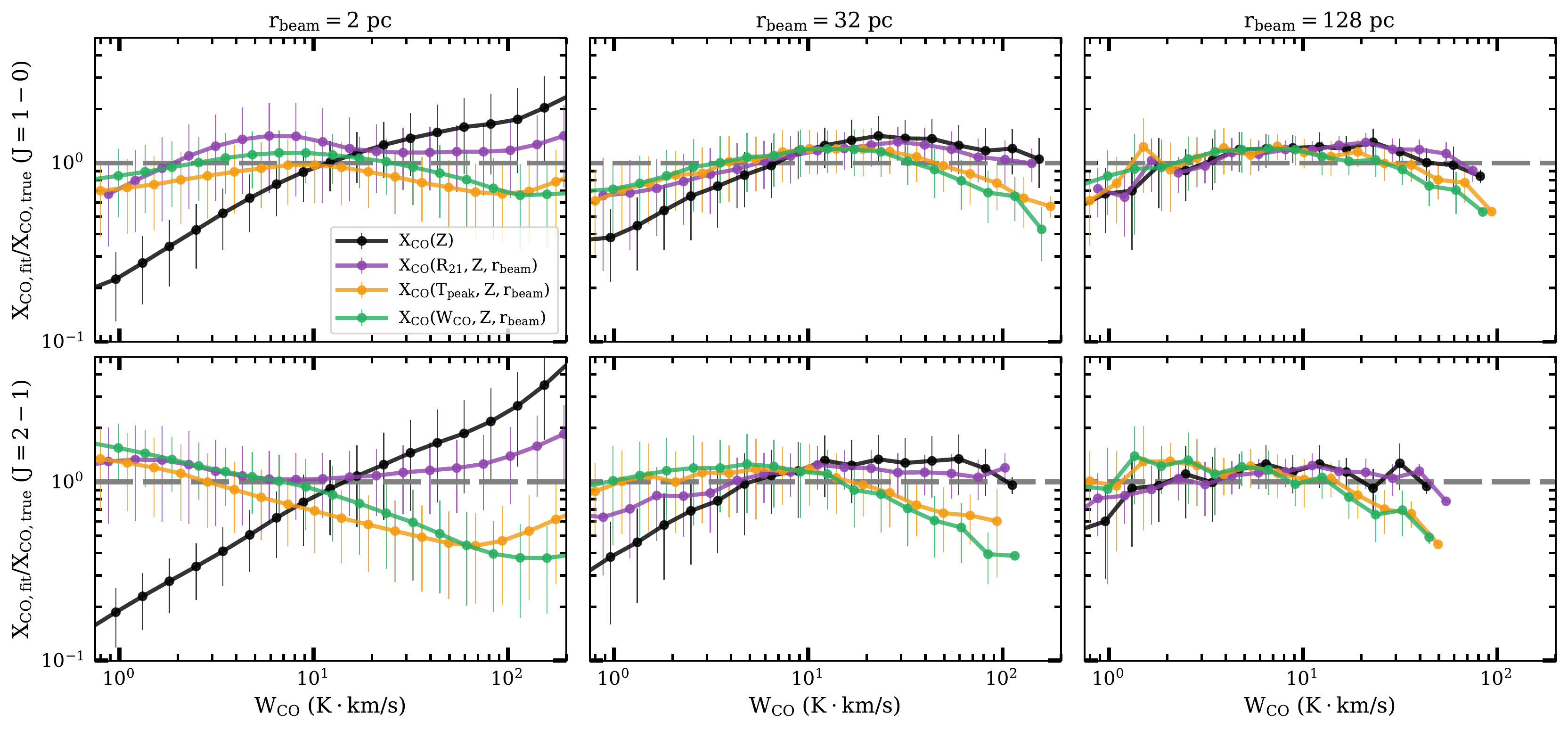}
\caption{Comparison of the $X_\CO$ fits to true values, binned by $W_\CO$.
  The symbols and error bars are the median value and semi-quartile
  range in each $W_\CO$ bin.
    The upper rows are for the
    CO(1--0) line and the lower rows the CO(2--1) line. The left,
    middle, and right columns are for synthetic observation with beam-sizes
    $r_\mr{beam} = 2$, $32$, and $128~\mr{pc}$. The black lines use
    the fit given in expressions
    1a/b of Table \ref{table:XCO_fit} that depends only
    on $Z$. In this case, $X_\CO$ is under-estimated in
    $\CO$-faint regions for small beams.
    The purple lines use the fit $X_\CO(R_{21}, Z, r_\mr{beam})$ that takes
    into account line ratios 
    (expressions 2a/b in Table \ref{table:XCO_fit}), which performs quite well
    overall.  The orange and green lines represent the fits
    $X_\CO(T_\mr{peak}, Z, r_\mr{beam})$ and $X_\CO(W_\CO, Z,r_\mr{beam})$ 
    (expressions 3a/b and 4a/b in Table \ref{table:XCO_fit}), which
    perform well in regions with low and moderate $W_\CO$,
    but under-estimate $X_\CO$ in the most
    CO-bright regions ($W_\CO \gtrsim 20~\mr{K\cdot km/s}$).
    \label{fig:XCO_fit_summary} }
\end{figure*}

As pointed out in Section \ref{section:XCO_physical},
there are significant systematic
variations in $X_\CO$ on small scales, correlated with the gas density and
excitation temperature. While these correlations reflect inherent
dependencies on physical conditions, neither the density nor the
excitation temperature is readily available from observations.
As a proxy, we identify direct observable quantities that
reflect physical conditions in a similar way, and use them to calibrate $X_\CO$
on small scales.

We consider the following observables: the metallicity $Z$,
the line ratio $R_{21}$, the peak antenna temperature $T_\mr{peak}$,
the integrated line intensity $W_\CO$, and the line width $\sigma_v$. 
We select the models R[2,4,8]-Z[05,1,2]
and R2B2-Z[05,1,2].  As discussed in \autoref{section:post-process} (see also \autoref{table:model}), these models have FUV radiation field that
matches the observed SFRs, which in R2 and R4 models requires a reduction
relative to the MHD model itself (the CRIR is scaled relative to the FUV).
The range of metallicity extends a factor of 2 above and below the solar
neighborhood.

Figures \ref{fig:XCO_corr_2}, \ref{fig:XCO_corr_32} and
\ref{fig:XCO_corr_128}  (see \autoref{sec:addendix}) show the values of
$X_\CO(1-0)$ and $X_\CO(2-1)$ for all $Z=1$ models as functions of
observables $R_{21}$, $T_\mr{peak}$, and $W_\CO$, for beam size $2$
pc, $32$ pc, and $128$ pc, respectively. 
For each observable and the range of beam sizes, we perform simple log-linear
fits using the least-squares method between the
observable and $X_\CO$, combining data from R2, R4, and R8 models.  Each data point
in the fitting represents a pixel in the synthetic observation, and
the fits are weighted by the area of the pixel.  We limit the fitting
to $\CO$-bright regions of $W_\CO > 0.75~\mr{K\cdot km/s}$.
The $X_\CO$-$T_\mr{peak}$ and $X_\CO$-$W_\CO$ relations are shallower at 
larger beam sizes due to beam-dilution. Therefore, we include an additional term $\log r_\mr{beam}$ 
in the power-law exponents of $T_\mr{peak}$ and $W_\CO$ to capture this effect. 
Due to beam-averaging, $X_\CO$ is roughly
constant when beam sizes are large, and we therefore limit the fitting
to $r_\mr{beam} \leq 128~\mr{pc}$. We also
tested $\sigma_v$, but found that it does not show any significant
correlation with $X_\CO$, as expected from Section
\ref{section:XCO_physical}; we therefore did not include it in the
final results. In addition, we experimented with fitting $\sigma_v$
together with other observables, and found no significant improvement
in the fit using the Bayesian information criteria. We fix the slopes 
for the $Z$ dependence ($X_\CO(1-0)\propto Z^{-0.8}$ and
$X_\CO(2-1)\propto Z^{-0.5}$), which were obtained from
fitting of median $X_\CO$ values in models with different metallicity
(see Section \ref{section:XCO_env} and Figure \ref{fig:XCO}).  We also
tried fitting the $X_\CO$ -- $Z$ relation using all pixels at the same
time, as we do for other variables, and obtained very similar slopes for $Z$. 

As can be seen from Figures \ref{fig:XCO_corr_2}, \ref{fig:XCO_corr_32} and
\ref{fig:XCO_corr_128}, the values of $X_\CO$ have large intrinsic
scatter at a given $R_{21}$, $T_\mr{peak}$, or $W_\CO$. This implies that other
hidden variables that are not directly observable, such as the detailed gas density,
temperature, and velocity structure along the line of sight, also influence $X_\CO$.
Although the relations between $X_\CO$ and the various observables are not true power-laws,
we find that the power-law fit we adopted already captures most of the systematic variations in the data. 
We find that the (absolute) difference between the fitted $X_\CO$ and the median values of $X_\CO$
in each bin is much smaller than the standard deviation of $X_\CO$ in each bin, 
except for the most CO-bright regions with $W_\CO \gtrsim 20~\mr{K\cdot km/s}$.
Even for $20~\mr{K\cdot km/s} \lesssim W_\CO \lesssim 200~\mr{K\cdot km/s}$,
the systematic errors from the power-law fit are still smaller than or comparable to 
the intrinsic scatter in $X_\CO$ (see also Figure \ref{fig:XCO_fit_summary}).

Table \ref{table:XCO_fit} summarizes the results of our fitting. In
expressions 1a/b, we provide our results for the 
relation with metallicity only from Figure \ref{fig:XCO}. 
Relations 2a/b, 3a/b, and 4a/b
give our calibrations for $X_\CO$ when the independent
variable is $R_{21}$, $T_\mr{peak}$, or $W_\CO$, respectively.
We note that $W_\CO$, $T_\mr{peak}$ and $R_{21}$ are
highly correlated, and therefore our fitted relationships should be
considered as set of alternative (rather than ``multiplicative'')
calibrations for $X_\CO$.

The fits for $X_\CO$ as functions of $R_{21}$, $T_\mr{peak}$, and $W_\CO$ are
included as dotted, dashed, and solid lines in Figures \ref{fig:XCO_corr_2},
\ref{fig:XCO_corr_32} and \ref{fig:XCO_corr_128}.
$R_{21}$, $T_\mr{peak}$ and $W_\CO$ all increase with
increasing gas density and excitation temperature, and thus negatively
correlate with $X_\CO$. At very high density $n\gtrsim 300~\mr{cm^{-3}}$ where
the optical depth for $\CO$ is very large, the
turn-over of $X_\CO$ in the left panel of Figure
\ref{fig:XCO_correlation_physical} is reflected in the
flattening of the binned $X_\CO$ values
near $W_\CO \approx 100~\mr{K\cdot km/s}$ and 
$T_\mr{peak} \approx 10~\mr{K}$.  $W_\CO \approx 100~\mr{K\cdot km/s}$ also
corresponds to the saturation level at $N_\Ht \gtrsim 5\times 10^{21}~\mr{cm^{-2}}$
in \autoref{fig:NH2_WCO}.
For the current physical conditions and
resolution in our simulations, most of the $\CO$ emission
comes from lower density regions where the trend in
Equation \ref{eq:XCO_approx1} is expected. $X_\CO$
decreases with increasing
$W_\CO$ and $T_\mr{peak}$ for the majority of the data points at high resolution.
Therefore, we simply use a single power-law fit. We do note, however, that our fits should not be applied to molecular cloud regions with 
$W_\CO \gtrsim 200~\mr{K\cdot km/s}$ where the lines are saturated. 

Comparing Figures \ref{fig:XCO_corr_2}, \ref{fig:XCO_corr_32} and
\ref{fig:XCO_corr_128}, it is apparent
that the scaling of $X_\CO$ with $T_\mr{peak}$ or $W_\CO$ 
is shallower at a larger $r_\mr{beam}$ due to beam-dilution. The slopes
for the $X_\CO$ fits are steeper for the (2--1) line, which 
traces regions with denser gas and higher excitation temperature than the
(1--0) line.

A comparison between all the $X_\CO$ fits and the original measurements,
binned by $W_\CO$,
is shown in Figure \ref{fig:XCO_fit_summary}. We present results separately for
$2$ pc, $32$ pc, and $128$  pc beams.
For smaller ($2$ pc or $32$ pc) beams,
the simple $X_\CO$ -- $Z$ relation is systematically biased: 
at low $W_\CO  \lesssim 10 ~\mr{K\cdot km/s}$, the relation 1a/b underestimates
the true $X_\CO$, while at high  $W_\CO \gtrsim  10 ~\mr{K\cdot km/s}$, the relation
1a/b slightly ($32$ pc) or significantly ($2$ pc)
overestimates the true $X_\CO$.  This can be
problematic when calculating masses of molecular clouds
with a large range of local physical conditions and brightness.
However, any of the three observables tested here
can help to correct this systematic bias.
$R_{21}$ performs the best across a large range of $W_\CO$, and the correlation
is insensitive to the beam-size. $W_\CO$ and $T_\mr{peak}$ perform well in
regions with low and moderate $W_\CO$, but
under-estimate $X_\CO$ when
$W_\CO \gtrsim 20~\mr{K\cdot km/s}$, with $T_\mr{peak}$ giving slightly better
results.

At $r_\mr{beam}\gtrsim 100~\mr{pc}$,
there is already significant averaging over varying density, temperature,
etc. within each beam, and we find that $X_\CO$ is consistent with having no correlation with $T_\mr{peak}$ or $W_\CO$. The $X_\CO$ dependencies on $Z$ and $R_{21}$, however, reflect the conditions for CO formation and excitation on all scales, and therefore do not suffer from beam dilution.  
In particular the $X_\CO$ relation with $R_{21}$  (2a/b in  \autoref{table:XCO_fit}) only has a very weak dependence on $r_\mr{beam}$ for the overall scaling 
at small beam sizes, and the dependence vanishes as beam sizes increase to $\gtrsim 100~\mr{pc}$. Therefore, for large beams, we recommend
using the simple $X_\CO$ -- $Z$ relation (1a/b in 
\autoref{table:XCO_fit}) if only a single line is available, or preferably the $X_\CO$ -- $R_{21}$ relation (2a/b) since this helps to capture the increase in excitation (and CO emission) in regions of higher mean density or where gas temperatures are enhanced by stronger heating.

\subsection{$\CO$-dark $\Ht$\label{section:fdark}}
\begin{figure}[htbp]
\centering
\includegraphics[width=\linewidth]{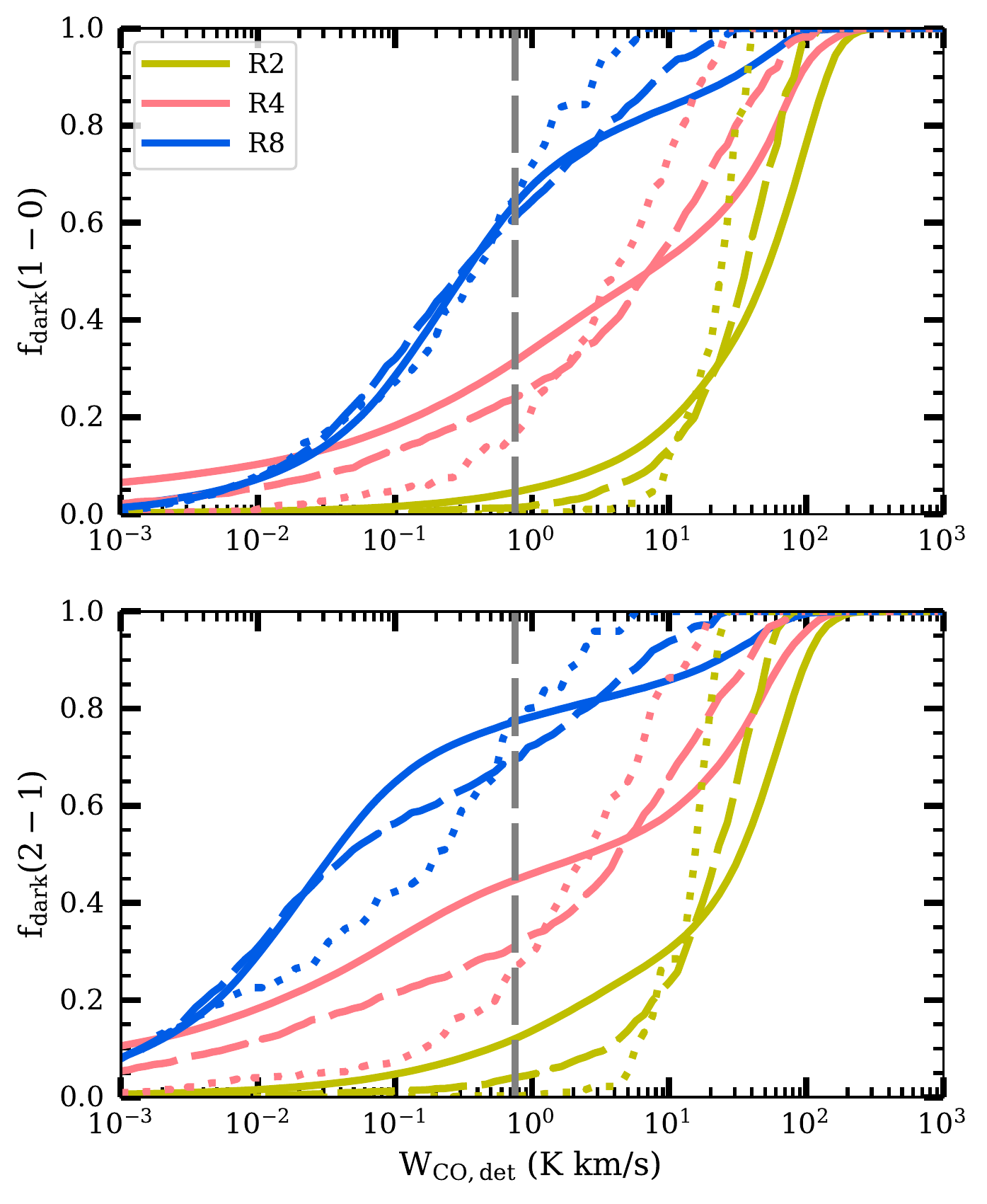}
    \caption{Fraction of CO-dark $\Ht$ as a function of the detection limit in
    the fiducial models (see legend), for both the (1--0) (top panel) and the (2--1) (bottom panel) lines.
    The different line styles show results from different beam-sizes $r_\mr{beam}=2~\mr{pc}$ (solid lines), $32~\mr{pc}$ (dashed lines)
    and $128~\mr{pc}$ (dotted lines).
    The vertical gray line shows the default detection limit in our studies.
    \label{fig:f_dark}
       }
\end{figure}
Finally, we investigate $f_\mr{dark}$, the CO-dark $\Ht$ fraction (defined in Equation \ref{eq:f_dark}). Figure
\ref{fig:f_dark} shows that in addition to the detection limit, $f_\mr{dark}$
also depends on the gas surface density, and to a lesser extent, the beam-size.
In the lower surface density R8 models, the clouds are fainter and smaller, and
thus fall more easily under the detection limit compared to
the brighter clouds in
R4 and R2 models. At the fiducial detection limit of $0.75~\mr{K\cdot km/s}$,
almost all the  $\Ht$  in the R2 model would be detected via $\CO$, 
while more than half of the $\Ht$ mass
remains CO-dark
in the R8 model  (see also Tables \ref{table:op32} and
\ref{table:op128}). 

\citet{Pety2013} analysed CO (1-0) line emission in M51 using different
observational data sets, and found that about
$50 \pm 10 \%$ of the emission is undetected at a resolution 
of 40 pc and sensitivity of $0.4~\mr{K\cdot km/s}$ ($1\sigma$). The average
surface density is about $30M_\odot \,{\rm pc}^{-2}$ in the regions they observed, similar to
that in our R4 models. We find that $f_\mr{dark}=30\%$ for the R4 models with
$r_\mr{beam}=32~\mr{pc}$ and $3\sigma$ detection limit of $1.2~\mr{K\cdot km/s}$, which can already
account for most of the missing emission in \citet{Pety2013}.

We also note from \autoref{table:op32} and \autoref{table:op128} that all models have a decrease in the fraction  of CO-dark gas at higher $Z$. However, especially for R2 and R4 models, $1-f_\mr{dark}$ varies little with $Z$. Since the majority of $\Ht$ is in CO-bright regions (for the range $Z=0.5-2$). This implies that in large-beam observations, the translation of  CO luminosity to $\Ht$ mass will depend on $Z$ mainly through the opacity of optically-thick lines (which affect the excitation temperature), as previously discussed (see Figure \ref{fig:Tgas_Texc_CRL} and related text).

\section{Conclusions}\label{section:conclusions}
In this paper, we use numerical simulations of the multiphase,
star-forming ISM in galactic disks to study the properties of the molecular
component and the $X_\CO$ conversion factor that is used to obtain $N_\Ht$ from
$W_\CO$. We extend the previous work of \citetalias{Gong2018}  based on
 simulations with solar neighborhood conditions
to a wide range of galactic environments. We
post-process 3D MHD simulations with chemistry and 
radiation transfer solvers
to produce synthetic maps of CO(1--0) and CO(2--1)
emission lines. We confirm numerical convergence of our results for
$X_\CO$ by varying the spatial resolution and box size. 

Our study investigates the dependencies on $X_\CO$ on large-scale environmental
parameters (metallicity, FUV radiation intensity, CRIR), 
local physical properties of the gas (density, excitation temperature),
and observables (CO brightness, peak temperature of the line,
line ratio), as well as averaging scale (beam size).
Our main findings are as follows:
\begin{enumerate}
    \item We successfully reproduce the relations between the $\CO$ peak
        brightness temperature $T_\mr{peak}$, the line width $\sigma_v$, and the
        brightness $W_\CO$ in the PHANGS survey of
        nearby galaxies 
        (Figure \ref{fig:WCO_PHANGS}), as well as
        the distribution of $R_{21}$, the CO (2--1) to (1--0) line
        ratio,  in the EMPIRE survey (Figure \ref{fig:C18_R4}). We also
        found a similar relation between $R_{21}$ and the FUV radiation field
        strength to that observed in M83 \citep{Koda2020}.
        This confirms that the
        molecular medium in our simulations is indeed a realistic
        representation of observed molecular clouds, for star-forming disk
        galaxies in the local Universe. 
      \item For varying metallicity  (relative to solar neighborhood)
        in the range of $Z=0.5-2$, we find $X_\CO \propto Z^{-0.8}$
        for the (1--0) line and $X_\CO \propto Z^{-0.5}$ for the (2--1) line (Figure
        \ref{fig:XCO}). This is consistent with
        observations of the Milky Way and nearby galaxies, and similar to
        results of other theoretical work (Figure
        \ref{fig:XCO_Z_obs}).  $X_\CO$ is reduced at higher $Z$ because
        of higher optical depth and higher $T_\mr{exc}$ at moderate density $n\approx 30-300\, \cm^{-3}$
        (Figure \ref{fig:Tgas_Texc_CRL};
        \autoref{eq:tau_LVG}  and \autoref{eq:Texc_approx}).
      \item $X_\CO$ decreases with increasing CRIR (Figure \ref{fig:XCO}),
        which increases heating and leads
        to higher $T_\mr{gas}$ and $T_\mr{exc}$ in the dense,
        shielded regions where CO forms 
      (Figure \ref{fig:Tgas_Texc_CRL}). $X_\CO$ is
        relatively insensitive to the FUV radiation field strength since higher
        FUV  increases $T_\mr{exc}$ only in weakly shielded regions with
        little CO, also partly
        compensating via a decreased optical depth. 
        The combined effect of CR and FUV would in principle lead
        to an anti-correlation between
        $X_\CO$ and the star formation rate for \textit{given} gas
        conditions (Figure \ref{fig:XCO}), although in practice star formation and gas conditions are correlated.
    \item On small scales, as the density increases, $X_\CO$ first decreases
        due to the increasing excitation temperature and then increases when 
        the emission is fully optically thick (Figures \ref{fig:Tgas_Texc} and
        \ref{fig:XCO_correlation_physical}). This is consistent with the
        theoretical expectations from 
        Equations \ref{eq:XCO_approx1} and \ref{eq:XCO_approx2}.
        Because the increase of $W_\mr{CO}$ with $N_\Ht$
        is steeper than linear at low
        $N_\Ht$ and flat at high $N_\Ht$ (Figure \ref{fig:NH2_WCO}), a constant
        $X_\CO$ is an underestimate at $N_\Ht \lesssim  2 \times 10^{21}~\mr{cm^{-2}}$ and
        an overestimate at $N_\Ht \gtrsim  2 \times 10^{21}~\mr{cm^{-2}}$.
    \item The direct observables $R_{21}$, $T_\mr{peak}$ and $W_\CO$ correlate
        with the gas density and the $\CO$ excitation temperature, and can be
        used to calibrate the systematic variations of $X_\CO$. We provide 
        fitting formulae for the calibration of
        $X_\CO$ in Table \ref{table:XCO_fit}. We show that
        using an $X_\CO$ that depends only on metallicity can introduce
        significant bias, especially at small beam-sizes (Figure
        \ref{fig:XCO_fit_summary}). For observations with $r_\mr{beam} \lesssim
        100 ~\mr{pc}$, we recommend using one of the observables $R_{21}$,
        $T_\mr{peak}$, or  $W_\CO$ to calibrate $X_\CO$. Among these choices,
        the calibration using $R_{21}$ performs the best in general, and can be used for large beams. The
        calibrations using $T_\mr{peak}$ and $W_\CO$ perform well at
        $W_\CO
        \lesssim 20~\mr{K\cdot km/s}$, and sightly over-estimate $X_\CO$ in
        higher brightness regions.
    \item The fraction of CO-dark $\Ht$ depends not only on sensitivity,
      but also on the gas surface density (and covariant environmental
      conditions) in galactic disks, and to a lesser
        extent, the beam-size. We provide an estimate of $f_\mr{dark}$ in
        Figure \ref{fig:f_dark}. The majority of 
        $\Ht$ is in CO-bright regions for higher surface density models at typical detection limits.  
\end{enumerate}
In the future, modeling of CO and calibration of
$X_\CO$ can be improved on two fronts. On the one
hand, galactic ISM simulations can be improved by including 
additional feedback mechanisms from star formation 
such as ionizing radiation and stellar winds, more accurate radiation transfer
from stellar clusters, injection and transport of CRs,
and covering a  larger range of parameter space 
beyond those in local disk galaxies. On the other hand, more accurate chemical
modelling can be achieved by coupling chemistry with radiation and
thermo-dynamics in the simulations. This will enable us to have a
fully
self-consistent model that follows the time-dependent
interactions between chemistry, 
metallicity evolution, radiation transfer, and gas dynamics. Currently, we are
working on improvements on both fronts within the TIGRESS framework.
Similar methods can also be used to
model the emission of other observable species,
such as $\mr{C^+}$, $\mr{CI}$, and
$\mr{HCO^+}$, which are valuable probes of  physical properties of different ISM
components.

\section{Acknowledgement}
We thank the anonymous referee for a constructive review, which helped to improve the overall clarity of this paper.
We thank Jiayi Sun and Adam Leroy for many helpful discussions and 
making the data from PHANGS available. We thank Diane Cormier for providing the
EMPIRE data. M. Gong
acknowledges support from Paola Caselli and the Max Planck 
Institute for Extraterrestrial Physics.
The work of E.C.O., C.-G.K., and J.-G.K was partially supported by
grants from NASA  (ATP award NNX17AG26G) and NSF (AARG award AST-1713949),
while C.-G.K. was additionally supported under Award No. CCA-528307
from the Simons Foundation. J.-G.K. acknowledges support from the Lyman Spitzer, Jr. Postdoctoral Fellowship at Princeton University.

This work used the open source MHD code Athena \citep{Stone2008, SG2009} and Athena++ \citep{Stone2020}, open source radiation transfer code RADMC-3D \citep{Dullemond2012}, and python packages Ipython \citep{Ipython}, numpy \citep{numpy}, scipy \citep{scipy}, matplotlib \citep{matplotlib}, astropy \citep{astropy1, astropy2}, yt \citep{yt} and lmfit \citep{lmfit}.

\bibliographystyle{apj}
\bibliography{apj-jour,all}

\appendix

\section{Additional Tables and Figures}
\label{sec:addendix}

\begin{figure*}[htbp]
\centering
\includegraphics[width=\linewidth]{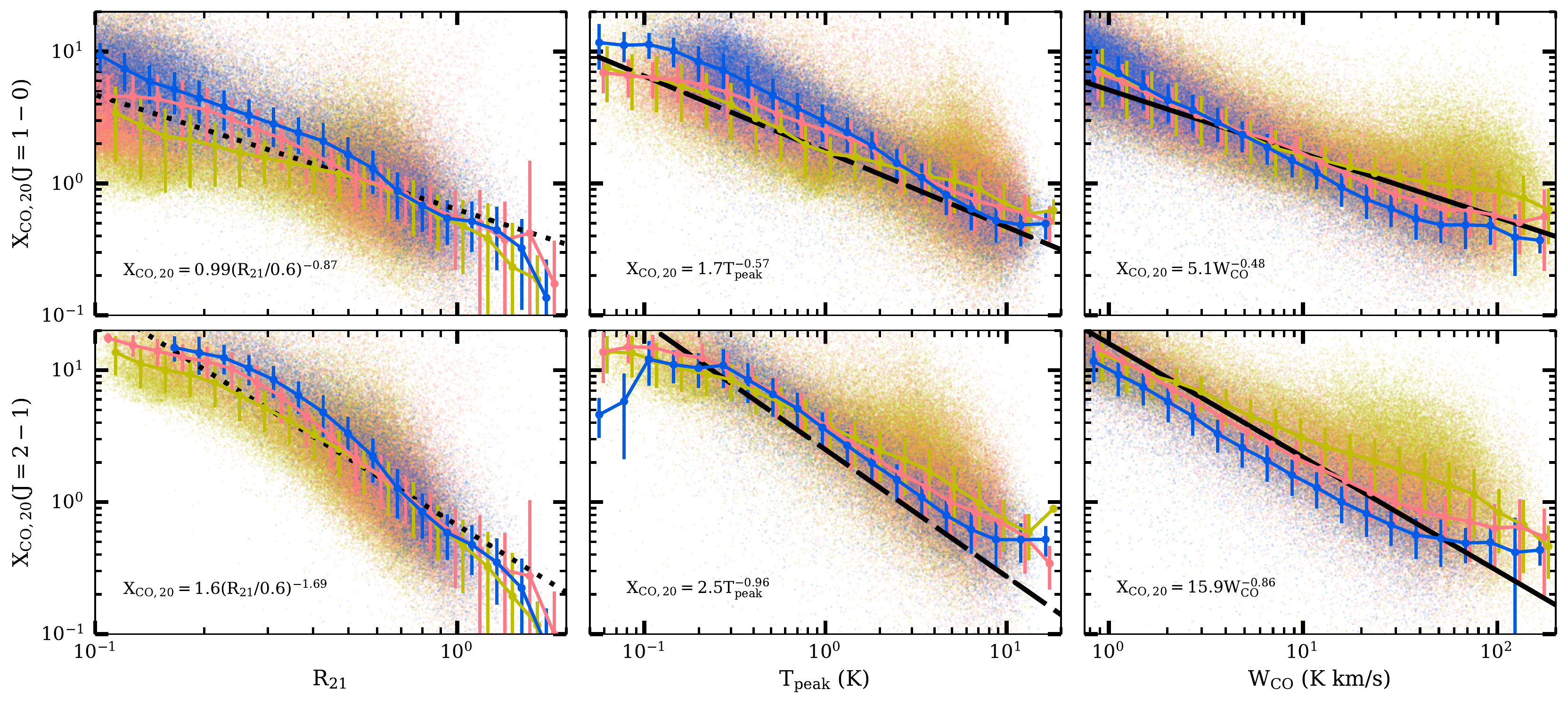}
    \caption{Correlation between $X_\CO$ and direct observables,
    for models R2-Z1 (yellow), R4-Z1 (red) and R8-Z1 (blue), at the native
    simulation beam-size of 2 pc. Only CO-bright regions with 
    $W_\CO > 0.75~\mr{K\cdot km/s}$ are shown and used for the fits.
    The binned median values and
    semi-quartile ranges are plotted over a the background of scatter points,
    each representing a pixel in the map.
    The black dotted, dashed, and solid lines are the fits 
    2a/b, 3a/b, 4a/b
    from Table \ref{table:XCO_fit}.
    \label{fig:XCO_corr_2}
    }
\end{figure*}

\begin{figure*}[htbp]
\centering
\includegraphics[width=\linewidth]{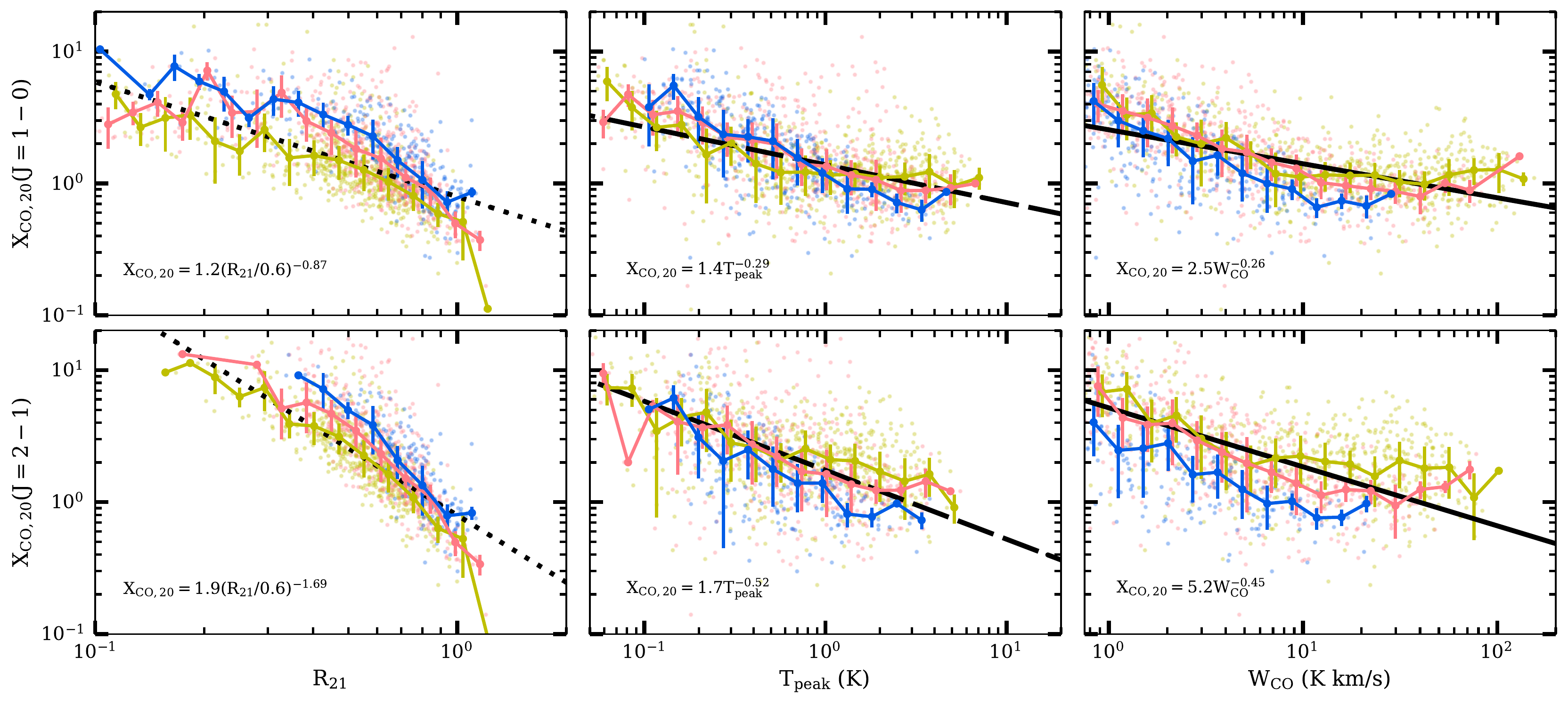}
    \caption{Same as Figure \ref{fig:XCO_corr_2}, but for a beam-size of 32 pc.
    \label{fig:XCO_corr_32}
       }
\end{figure*}

\begin{figure*}[htbp]
\centering
\includegraphics[width=\linewidth]{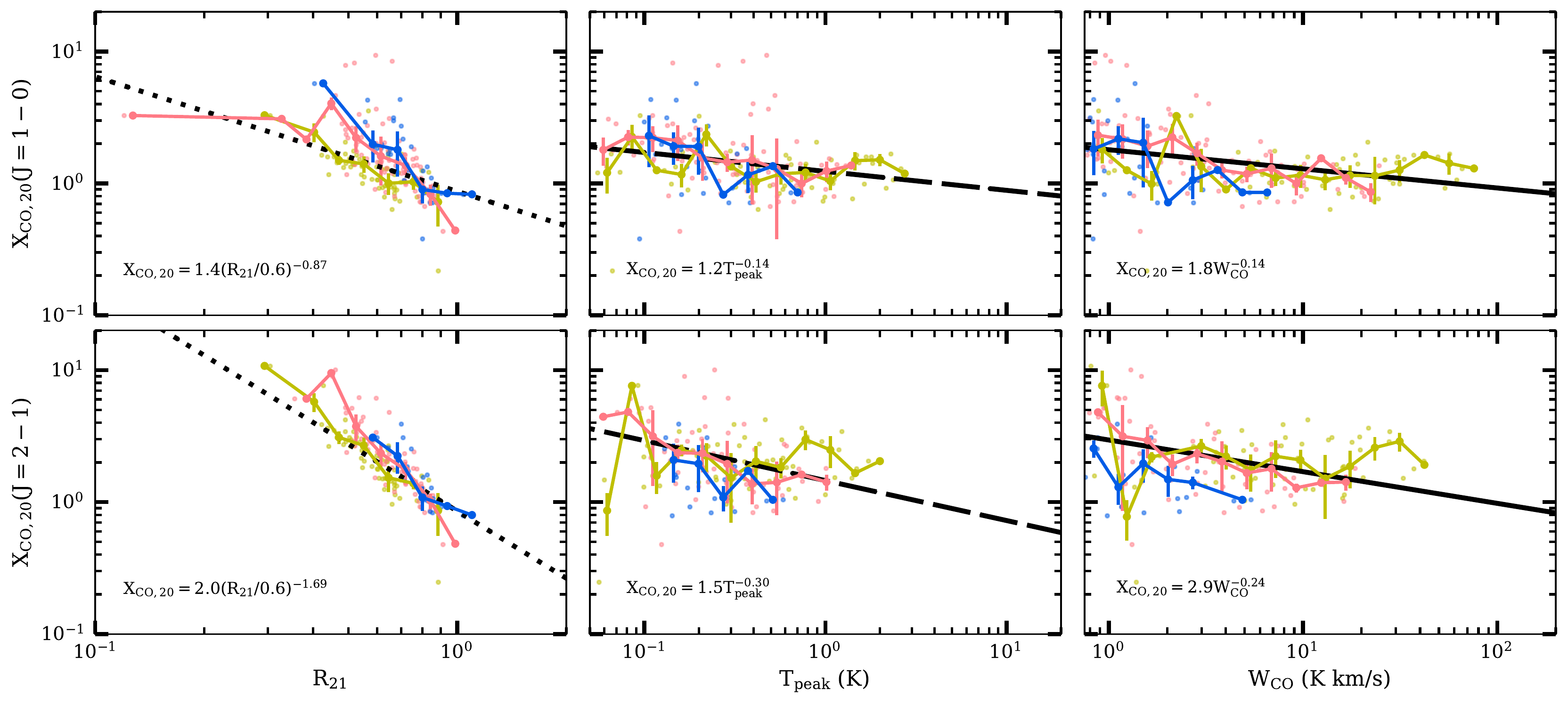}
    \caption{Same as Figure \ref{fig:XCO_corr_2}, but for a beam-size of 128 pc.
    \label{fig:XCO_corr_128}
       }
\end{figure*}

Additional Tables \ref{table:op32} and \ref{table:op128} are included, detailing the overall  properties of the simulations. Additional Figures \ref{fig:XCO_corr_2}, \ref{fig:XCO_corr_32} and \ref{fig:XCO_corr_128} are included to show the fits for $X_\CO$. 
\begin{table*}[htbp]
    \caption{Overall properties of simulation with 32 pc beam in $\CO$-bright
    regions\tablenotemark{a}}
    \label{table:op32}
    \centering
    \setlength\tabcolsep{2.4pt}
    \begin{tabular}{l c ccccc c ccccc}
        \tableline
        \tableline
        \multirow{2}{*}{Model} &\multicolumn{6}{c}{$\CO(J=1-0)$}
         &\multicolumn{6}{c}{$\CO(J=2-1)$}\\\cmidrule(lr){2-7}\cmidrule(lr){8-13}
        &$N_\mr{H_2,20}$ &$X_\mr{CO,20}$ &$W_\mr{CO}$ &$\sigma_v$ &$T_\mr{peak}$ &$f_\mr{dark}$
        &$N_\mr{H_2,20}$ &$X_\mr{CO,20}$ &$W_\mr{CO}$ &$\sigma_v$ &$T_\mr{peak}$ &$f_\mr{dark}$\\
        \tableline
        \multicolumn{13}{l}{Physics model:}\\
        R2-Z1CR10L10  &7.36(6.6) &0.67(0.2) &11.09(11.8) &6.01(1.5) &0.94(0.9) &0.013
                      &9.19(7.1) &0.89(0.4) &9.17(11.6) &5.69(1.4) &0.82(0.9) &0.026\\
        R2-Z1L10      &14.37(10.0) &1.04(0.4) &13.00(13.0) &6.12(1.6) &1.07(1.0) &0.024
                      &16.05(10.1) &1.42(0.7) &10.84(10.4) &5.78(1.5) &0.89(0.8) &0.042\\
        R2-Z1CR10     &8.47(7.0) &0.77(0.2) &11.13(12.0) &6.38(1.5) &0.88(0.9) &0.015
                      &11.06(7.8) &1.04(0.5) &10.02(11.6) &5.82(1.4) &0.82(0.9) &0.037\\
        \textbf{R2-Z1} &16.68(11.4) &1.13(0.4) &15.81(13.8) &6.88(1.6) &1.17(0.9) &0.013
                      &19.55(11.6) &1.81(0.9) &11.98(9.5) &6.41(1.6) &0.86(0.6) &0.040\\
        R2-Z1L01      &18.51(12.5) &1.14(0.3) &18.49(15.0) &7.54(1.5) &1.17(0.8) &0.010
                      &21.66(12.6) &2.03(0.8) &12.16(9.1) &6.86(1.6) &0.82(0.6) &0.037\\
        R2-Z1CR01     &25.70(13.0) &1.70(0.8) &15.70(12.6) &6.52(1.7) &1.19(0.8) &0.038
                      &26.81(13.6) &2.94(1.4) &9.70(6.7) &6.34(1.6) &0.71(0.4)  &0.063\\
        R2-Z1CR01L01  &30.69(15.6) &1.81(0.7) &19.40(13.1) &7.66(1.5) &1.15(0.7) &0.027
                      &33.28(15.2) &4.52(1.6) &8.70(5.0) &7.16(1.6) &0.52(0.3) &0.058\\
        R2-Z05        &18.67(10.6) &2.11(1.2) &8.20(9.0) &5.55(1.4) &0.78(0.7) &0.074
                      &21.22(11.4) &2.57(1.3) &7.31(7.1) &5.18(1.2) &0.70(0.5) &0.130\\
        R2-Z2         &18.33(13.3) &0.68(0.2) &29.04(19.1) &8.20(1.5) &1.59(1.0) &0.004
                      &20.73(13.4) &1.43(0.5) &17.38(11.6) &7.78(1.6) &0.99(0.6) &0.014\\
        R4-Z1CR10L10  &4.05(3.4) &0.89(0.4) &3.88(4.1) &2.98(1.1) &0.63(0.5) &0.136
                      &5.20(4.2) &0.99(0.6) &4.14(4.3) &3.01(1.0) &0.63(0.5) &0.186\\
        R4-Z1L10      &8.38(5.6) &1.29(0.8) &5.68(4.8) &2.93(1.0) &0.86(0.6) &0.234
                      &8.96(5.7) &1.46(1.0) &4.63(4.2) &2.92(1.1) &0.72(0.5) &0.253\\
        R4-Z1CR10     &4.28(3.5) &0.97(0.4) &3.98(4.6) &3.32(1.1) &0.60(0.6) &0.167
                      &6.16(4.4) &1.15(0.7) &4.68(4.9) &3.13(1.1) &0.67(0.5) &0.235\\
        \textbf{R4-Z1} &8.70(5.1) &1.65(0.9) &4.93(5.2) &3.52(1.2) &0.71(0.6) &0.242
                      &10.66(6.2) &1.99(1.3) &4.48(4.1) &3.25(1.1) &0.62(0.4) &0.310\\
        R4-Z1L01      &7.69(4.7) &1.61(0.9) &4.52(4.5) &4.32(1.3) &0.58(0.5) &0.211
                      &11.02(5.7) &2.37(1.5) &4.15(3.6) &3.76(1.2) &0.53(0.4) &0.340\\
        R4-Z1CR01     &14.31(6.4) &2.37(1.6) &5.08(4.9) &3.09(1.1) &0.72(0.5) &0.372
                      &15.55(7.0) &3.33(2.2) &3.85(3.2) &3.16(1.2) &0.53(0.3) &0.410\\
        R4-Z1CR01L01  &15.34(6.0) &2.80(1.9) &4.97(4.6) &3.93(1.2) &0.63(0.5) &0.370
                      &17.99(6.8) &5.09(2.7) &3.13(2.3) &3.76(1.2) &0.38(0.2) &0.470\\
        R4-Z05        &12.86(6.1) &2.90(1.9) &3.52(3.4) &2.86(0.9) &0.59(0.4) &0.384
                      &13.66(6.6) &3.30(2.0) &3.35(2.9) &3.00(1.0) &0.49(0.3) &0.441\\
        R4-Z2         &6.40(4.7) &0.91(0.5) &7.00(6.5) &4.36(1.5) &0.77(0.7) &0.134
                      &8.57(5.1) &1.57(1.0) &5.23(4.9) &4.06(1.3) &0.64(0.5) &0.223\\
        \textbf{R8-Z1} &4.83(2.2) &1.70(1.1) &2.42(1.9) &2.01(0.5) &0.48(0.3) &0.612
                      &5.38(2.6) &1.67(0.9) &2.64(2.2) &2.00(0.4) &0.47(0.3) &0.696\\
        R8-Z05        &6.35(2.3) &2.76(1.1) &2.11(1.5) &1.88(0.4) &0.45(0.3) &0.840
                      &6.32(2.6) &2.31(1.1) &1.95(1.7) &1.93(0.4) &0.42(0.2) &0.855\\
        R8-Z2         &3.28(1.7) &0.92(0.5) &3.08(2.9) &2.31(0.6) &0.54(0.4) &0.306
                      &4.26(2.0) &1.13(0.7) &3.18(2.6) &2.23(0.5) &0.52(0.4) &0.459\\
        \tableline
        \multicolumn{13}{l}{Convergence of simulation box-size:}\\
        R2B2-Z1       &20.65(17.4) &1.26(0.5) &17.78(17.8) &7.26(1.4) &1.16(1.0) &0.024
                      &23.31(18.8) &2.25(1.1) &12.17(10.9) &6.81(1.5) &0.82(0.6) &0.044\\
        R2B2-Z05      &25.22(19.4) &2.21(1.1) &11.94(13.0) &6.02(1.4) &0.98(0.9) &0.067
                      &28.41(19.6) &3.28(1.7) &9.37(9.3) &5.70(1.4) &0.74(0.6) &0.091\\
        R2B2-Z2       &16.55(17.1) &0.83(0.3) &25.97(22.1) &8.64(1.4) &1.39(0.9) &0.011
                      &21.49(18.3) &1.65(0.8) &15.04(11.1) &8.12(1.5) &0.84(0.5) &0.023 \\
        \tableline
        \multicolumn{13}{l}{Convergence of numerical resolution}\\
        R2N2-Z1       &19.64(11.1) &1.18(0.3) &17.89(11.0) &6.89(1.2) &1.15(0.8) &0.007
                      &20.63(10.9) &1.83(0.6) &11.70(7.3) &6.65(1.4) &0.79(0.5) &0.015\\
        R2N2-Z05      &18.68(9.3) &2.48(1.2) &6.64(5.7) &5.43(1.3) &0.64(0.5) &0.068
                      &20.88(10.8) &3.04(1.5) &5.87(4.7) &5.31(1.4) &0.53(0.4) &0.114\\
        R2N2-Z2       &24.05(13.4) &0.74(0.1) &38.09(16.0) &8.49(0.9) &1.82(0.8) &0.001
                      &24.24(13.3) &1.42(0.4) &19.16(9.4) &8.10(1.1) &1.05(0.5) &0.002\\
        \tableline
        \tableline
    \end{tabular}
    \tablenotetext{1}{All variables are calculated from $\CO$-bright regions,
    which are defined as beams with $W_\CO > 0.75~\mr{K\cdot km/s}$. The median
    values of the variables in all beams are shown as the main number,
    with the semi-quartile range shown in the following brackets. $N_\mr{H_2,20} =
    N_\Ht/(10^{20}~\mr{cm^{-2}})$. $X_\mr{CO,20}=X_\CO/(10^{20}~\mr{cm^{-2}
    K^{-1} km^{-1} s})$. $W_\CO$ is in units of $\mr{K~km/s}$. $\sigma_v$ is
    the velocity dispersion of the $\CO$ line profile. $T_\mr{peak}$ is the
    peak brightness temperature of the $\CO$ line profile. $f_\mr{dark}$ is the
    fraction of $\CO$-dark $\Ht$. The fiducial model names are highlighted in bold.
    }
\end{table*}

\begin{table*}[htbp]
    \caption{Overall properties of simulation with 128 pc beam in $\CO$-bright
    regions\tablenotemark{a}}
    \label{table:op128}
    \centering
    \setlength\tabcolsep{2.4pt}
    \begin{tabular}{l c ccccc c ccccc}
        \tableline
        \tableline
        \multirow{2}{*}{Model} &\multicolumn{6}{c}{$\CO(J=1-0)$}
         &\multicolumn{6}{c}{$\CO(J=2-1)$}\\\cmidrule(lr){2-7}\cmidrule(lr){8-13}
        &$N_\mr{H_2,20}$ &$X_\mr{CO,20}$ &$W_\mr{CO}$ &$\sigma_v$ &$T_\mr{peak}$ &$f_\mr{dark}$
        &$N_\mr{H_2,20}$ &$X_\mr{CO,20}$ &$W_\mr{CO}$ &$\sigma_v$ &$T_\mr{peak}$ &$f_\mr{dark}$\\
        \tableline
        \multicolumn{13}{l}{Physics model:}\\
        R2-Z1CR10L10 &7.26(5.0) &0.60(0.1) &10.32(7.3) &7.87(1.4) &0.58(0.3) &0.001
                     &7.32(5.0) &0.79(0.2) &7.70(6.1) &7.90(1.4) &0.48(0.3) &0.002\\
        R2-Z1L10     &12.48(7.4) &0.96(0.2) &10.55(7.4) &7.79(1.3) &0.63(0.3) &0.001
                     &12.48(7.4) &1.30(0.2) &7.81(5.4) &7.95(1.3) &0.47(0.2) &0.001\\
        R2-Z1CR10    &9.17(5.7) &0.68(0.1) &12.07(7.4) &8.05(1.2) &0.62(0.3) &0.001
                     &9.52(5.8) &0.86(0.2) &8.79(5.7) &7.97(1.2) &0.50(0.3) &0.003\\
        \textbf{R2-Z1} &17.00(8.9) &1.04(0.1) &16.03(8.4) &8.36(1.3) &0.75(0.3) &0.003
                     &17.00(8.9) &1.63(0.3) &9.07(5.6) &8.38(1.2) &0.53(0.2) &0.003\\
        R2-Z1L01     &19.31(9.4) &1.08(0.2) &19.28(9.3) &8.96(1.0) &0.87(0.4) &0.000
                     &19.99(9.5) &1.91(0.4) &10.49(5.1) &8.68(1.1) &0.55(0.2) &0.004\\
        R2-Z1CR01    &22.79(10.0) &1.66(0.2) &12.96(6.9) &8.15(1.3) &0.68(0.3) &0.004
                     &22.81(9.7) &2.86(0.4) &7.30(3.6) &8.27(1.3) &0.43(0.2) &0.007\\
        R2-Z1CR01L01 &28.69(12.4) &1.77(0.3) &16.60(8.1) &9.23(0.9) &0.78(0.3) &0.000
                     &30.79(11.2) &4.26(0.8) &7.16(2.9) &8.91(1.0) &0.34(0.1) &0.008\\
        R2-Z05       &13.22(7.5) &1.80(0.5) &5.74(4.2) &7.19(1.2) &0.32(0.2) &0.004
                     &13.60(6.7) &2.49(0.9) &5.10(2.8) &7.47(1.2) &0.29(0.1) &0.015\\
        R2-Z2        &20.53(9.9) &0.68(0.1) &29.59(13.9) &9.76(0.9) &1.23(0.4) &0.000
                     &20.53(9.9) &1.21(0.3) &16.31(6.5) &9.50(0.9) &0.69(0.2) &0.000\\
        R4-Z1CR10L10 &2.52(1.8) &0.80(0.2) &2.40(2.2) &4.20(1.5) &0.24(0.1) &0.161
                     &2.88(2.3) &0.93(0.3) &2.20(2.7) &4.88(1.6) &0.24(0.1) &0.205\\
        R4-Z1L10     &3.80(2.6) &1.35(0.4) &2.41(2.1) &4.54(1.5) &0.25(0.2) &0.205
                     &3.88(2.9) &1.56(0.6) &2.28(2.2) &5.00(1.6) &0.20(0.1) &0.242\\
        R4-Z1CR10    &2.54(1.7) &0.86(0.2) &2.68(2.3) &4.69(1.5) &0.29(0.2) &0.154
                     &3.35(1.9) &1.12(0.4) &2.64(2.6) &4.40(1.7) &0.26(0.2) &0.216\\
        \textbf{R4-Z1} &5.03(3.0) &1.48(0.5) &2.90(2.3) &4.70(1.5) &0.31(0.2) &0.167
                     &5.89(3.2) &2.03(0.8) &2.58(1.8) &4.41(1.5) &0.22(0.1) &0.268\\
        R4-Z1L01     &6.35(3.2) &1.36(0.5) &3.78(3.0) &5.87(1.4) &0.37(0.2) &0.148
                     &7.22(2.9) &2.41(0.9) &2.22(1.8) &5.17(1.5) &0.24(0.1) &0.268\\
        R4-Z1CR01    &8.27(4.1) &2.46(1.0) &2.65(2.0) &3.97(1.4) &0.27(0.1) &0.275
                     &8.87(4.6) &3.59(1.3) &2.12(1.5) &4.86(1.5) &0.20(0.1) &0.362\\
        R4-Z1CR01L01 &11.09(3.8) &2.98(1.5) &3.00(2.2) &5.28(1.4) &0.32(0.1) &0.233
                     &13.07(4.2) &5.70(1.9) &1.91(1.2) &5.33(1.5) &0.16(0.1) &0.394\\
        R4-Z05       &6.93(3.6) &2.33(1.0) &2.51(1.9) &5.15(1.5) &0.20(0.1) &0.420
                     &7.15(4.3) &2.59(1.0) &3.05(1.2) &5.55(1.4) &0.21(0.1) &0.513\\
        R4-Z2        &4.95(3.2) &0.86(0.3) &5.12(4.2) &5.76(1.4) &0.43(0.3) &0.071
                     &6.02(3.3) &1.53(0.7) &3.04(2.5) &5.63(1.3) &0.29(0.2) &0.141\\
        \textbf{R8-Z1} &2.48(1.4) &1.51(0.6) &1.31(0.9) &2.93(0.4) &0.20(0.1) &0.652
                     &2.71(1.4) &1.49(0.6) &1.64(0.6) &2.90(0.4) &0.22(0.1) &0.775\\
        R8-Z05       &3.63(0.9) &2.31(0.2) &1.61(0.5) &3.42(0.7) &0.18(0.0) &0.900
                     &3.74(0.1) &2.58(0.2) &1.58(0.1) &3.65(0.6) &0.16(0.0) &0.923\\
        R8-Z2        &2.07(0.9) &0.78(0.4) &2.29(1.2) &3.08(0.7) &0.32(0.2) &0.275
                     &2.24(1.3) &1.07(0.5) &1.75(0.8) &3.16(0.7) &0.25(0.1) &0.400\\        
        \tableline
        \multicolumn{13}{l}{Convergence of simulation box-size:}\\
        R2B2-Z1      &13.51(11.1) &1.27(0.3) &13.04(9.0) &9.07(0.9) &0.57(0.4) &0.009
                     &14.59(11.1) &2.05(0.6) &8.05(5.4) &9.00(0.9) &0.36(0.2) &0.015\\
        R2B2-Z05     &13.67(9.9) &2.03(0.4) &6.57(5.1) &8.21(0.9) &0.32(0.2) &0.029
                     &13.89(11.7) &2.89(1.0) &4.82(3.2) &8.17(0.9) &0.24(0.2) &0.036\\
        R2B2-Z2      &13.94(12.2) &0.81(0.2) &19.26(13.3) &9.90(0.7) &0.76(0.4) &0.004
                     &14.83(12.9) &1.54(0.6) &10.80(7.0) &9.73(0.8) &0.44(0.2) &0.007\\
        \tableline
        \multicolumn{13}{l}{Convergence of numerical resolution}\\
        R2N2-Z1      &21.36(5.1) &1.13(0.1) &19.77(4.2) &8.62(0.5) &0.92(0.2) &0.000
                     &21.36(5.1) &1.83(0.2) &12.71(2.4) &8.59(0.6) &0.59(0.1) &0.000\\
        R2N2-Z05     &16.78(4.7) &2.35(0.3) &7.80(3.0) &7.53(0.5) &0.43(0.2) &0.000
                     &16.78(4.7) &3.33(0.5) &5.66(2.1) &7.55(0.5) &0.31(0.1) &0.000\\
        R2N2-Z2      &26.28(5.6) &0.80(0.1) &35.90(4.4) &9.43(0.3) &1.46(0.1) &0.000
                     &26.28(5.6) &1.42(0.1) &19.67(2.4) &9.31(0.4) &0.82(0.1) &0.000\\
        \tableline
        \tableline
    \end{tabular}
    \tablenotetext{1}{Same as Table \ref{table:op32} but with a 128 pc beam.
    }
\end{table*}

\end{document}